\documentclass[journal]{IEEEtran}
\ifCLASSINFOpdf
\else
\fi
\usepackage{nomencl}
\usepackage{amssymb}
\usepackage{float}
\usepackage{graphicx}
\usepackage{subfigure}
\usepackage{indentfirst}
\usepackage{etoolbox}
\usepackage{cite}
\usepackage{amsmath,amssymb,amsfonts}
\usepackage{algorithmic}
\usepackage{mathtools}
\usepackage{textcomp}
\usepackage{xcolor}
\usepackage{multirow}
\usepackage{makecell}
\usepackage{booktabs}
\usepackage{tipa}
\usepackage{dsfont}
\usepackage{textcomp}
\usepackage{threeparttable}
\usepackage{threeparttable}
\usepackage{bm}
\usepackage{booktabs}
\usepackage{siunitx}
\usepackage{makecell}
\usepackage{verbatim}
\usepackage{graphicx}
\usepackage{textcomp}
\usepackage{pifont}
\usepackage{newtxtext}
\makenomenclature

\begin{document}
\title{Tri-Level Stochastic-Robust Co-Planning of Distribution Networks and Renewable Charging Stations With an Adaptive iC\&CG Algorithm}
\author{Yongheng~Wang, Xiemin~Mo, Tao~Liu*
\thanks{This work was supported by the National Natural Science Foundation of China through Project No. 62173287 and the Research Grants Council of the Hong Kong Special Administrative Region under the Early Career Scheme through Project No. 27206021.}
\thanks{Y. Wang, X. Mo and T. Liu (corresponding author: taoliu@eee.hku.hk) are with the Department of Electrical and Electronic Engineering, The University of Hong Kong, Hong Kong SAR.}
}

\maketitle
\begin{abstract}
Renewable charging stations (RCSs) that co-locate electric-vehicle (EV) charging with distributed generation (DG) can raise renewable utilization and improve distribution-network (DN) efficiency, yet their variability and the siting-dependent charging demand can overload feeders if placed poorly. This paper proposes a tri-level, two-stage stochastic--robust optimization (SRO) co-planning framework that jointly determines RCS siting and DN expansion while accounting for transportation flows and population density. The model distinguishes two uncertainty classes: (i) \emph{decision-dependent uncertainty} (DDU), under which EV charging loads vary with RCS siting; and (ii) \emph{decision-independent uncertainty} (DIU), under which load fluctuations and renewable-generation variability do not depend on the RCS locations or the DN topology. At the upper level, the framework selects RCS sites and DN expansions. At the middle level, EV routing and charging are dispatched given the RCS siting to produce charging loads DDU. At the lower level, DN operation minimizes annualized loss costs under the worst-case DIU, reformulated via Karush--Kuhn--Tucker (KKT) conditions. To solve the resulting problem efficiently, we develop an adaptive inexact column-and-constraint generation (A-iC\&CG) algorithm and prove finite-iteration convergence. Case studies on a 47-node DN coupled with a 68-hub transportation network in Shenzhen, China, show that A-iC\&CG outperforms benchmark algorithms and that PV--EV hybrid stations are cost-optimal, with RCS siting concentrated near substations and high-flow hubs.
\end{abstract}
\vspace{-0.2em}

\vspace{-1em}

\IEEEpeerreviewmaketitle
\renewcommand\nomgroup[1]{%
  \item[\bfseries
  \ifstrequal{#1}{A}{Abbreviations}{%
  \ifstrequal{#1}{B}{Sets and Indices}{%
  \ifstrequal{#1}{C}{Parameters}{%
  \ifstrequal{#1}{D}{Variables}{}}}}%
]}
\nomenclature[A,01]{EV}{Electric vehicle}
\nomenclature[A,02]{PV}{Photovoltaic}
\nomenclature[A,03]{ESS}{Energy storage system}
\nomenclature[A,04]{RCS}{Renewable charging station}
\nomenclature[A,05]{DN}{Distribution network}
\nomenclature[A,06]{RO}{Robust optimization}
\nomenclature[A,07]{SO}{Stochastic optimization}
\nomenclature[A,08]{DRO}{Distributionally robust optimization}
\nomenclature[A,09]{SRO}{Stochastic robust optimization}
\nomenclature[A,10]{DDU}{Decision-dependent uncertainty}
\nomenclature[A,11]{DIU}{Decision-independent uncertainty}
\nomenclature[A,12]{C\&CG}{Column-and-constraint generation}
\nomenclature[A,13]{NC\&CG}{Nested column-and-constraint generation}
\nomenclature[A,14]{iC\&CG}{Inexact column-and-constraint generation}
\nomenclature[A,15]{A-iC\&CG}{Adaptive inexact column-and-constraint generation}
\nomenclature[B,01]{\(\mathbb{E}/u\)}{Set/index of EVs}
\nomenclature[B,02]{\(\mathbb{N}/i\)}{Set/index of nodes}
\nomenclature[B,03]{\(\mathbb{T}/t\)}{Set/index of time intervals}
\nomenclature[B,04]{\(\mathbb{L}/(i,j)\)}{Set/index of distribution lines}
\nomenclature[B,05]{\(\mathbb{H}/k\)}{Set/index of transportation hubs}
\nomenclature[B,06]{\(\mathbb{C}\)}{Set of charging time periods for EVs}
\nomenclature[C,01]{\(\varphi_{ij}\)}{Construction and maintenance cost of line \(ij\)}
\nomenclature[C,02]{\(\varphi_{rcs}\)}{Construction and maintenance cost of RCS}
\nomenclature[C,03]{\(D_{i}\)}{Fictitious load demand at node \(i\)}
\nomenclature[C,04]{\(m\)}{Number of RCS planned in residential, office, commercial, and industrial areas}
\nomenclature[C,05]{\(m_{k,u}^{in}\)}{Entry in the \(k\)-th row and \(u\)-th column of the spatio-temporal matrix indicating that EV \(u\) is at hub \(k\) (requesting a charge)}
\nomenclature[C,06]{\(R_{kk}\)}{Transportation information matrix storing distances among the \(k\) hubs}
\nomenclature[C,07]{\(c_{t}^{tou}\)}{Time-of-use electricity price at time $t$}
\nomenclature[C,07]{\(c^{ev}\)}{Travel cost factor of EVs}
\nomenclature[C,08]{\(\overline{s_{ij}}\)}{Apparent power capacity of line \(ij\)}
\nomenclature[C,09]{\(\underline{n_{k}},\,\overline{n_{k}}\)}{Minimum and maximum number of EVs at hub \(k\)}
\nomenclature[C,10]{\(e_{u,0},\,e_{u}\)}{Initial and target energy state of EV \(u\)}
\nomenclature[C,11]{\(\underline{e_u},\,\overline{e_u}\)}{Minimum and maximum energy capacity of EV \(u\)}
\nomenclature[C,12]{\(\underline{p_u},\,\overline{p_u}\)}{Minimum and maximum charging power of EV \(u\)}
\nomenclature[C,13]{\(\eta_{ch,k}^{ess},\,\eta_{dis,k}^{ess}\)}{Charging and discharging efficiency of the ESS at hub \(k\)}
\nomenclature[C,14]{\(\underline{p_{k}^{ch}},\,\overline{p_{k}^{ch}}\)}{Minimum and maximum charging power of the ESS at hub \(k\)}
\nomenclature[C,15]{\(\underline{p_{k}^{dis}},\,\overline{p_{k}^{dis}}\)}{Minimum and maximum discharging power of the ESS at hub \(k\)}
\nomenclature[C,16]{\(\underline{e_{k}^{ess}},\,\overline{e_{k}^{ess}}\)}{Minimum and maximum energy capacity of the ESS at hub \(k\)}
\nomenclature[C,17]{\(r_{ij},\,x_{ij}\)}{Resistance and reactance of line \(ij\)}
\nomenclature[C,18]{\(\underline{v},\,\overline{v}\)}{Minimum and maximum voltage magnitude}
\nomenclature[C,19]{\(\underline{p^{sub}},\,\overline{p^{sub}}\)}{Minimum and maximum active power output of the substation}
\nomenclature[C,20]{\(\underline{q^{sub}},\,\overline{q^{sub}}\)}{Minimum and maximum reactive power output of the substation}
\nomenclature[C,21]{\(\underline{p^{pv}},\,\overline{p^{pv}}\)}{Minimum and maximum active power output of the PV unit at node \(i\) and time \(t\)}
\nomenclature[C,22]{\(\underline{p_{i,t}^{load}},\,\overline{p_{i,t}^{load}}\)}{Minimum and maximum active power load at node \(i\) and time \(t\)}
\nomenclature[C,23]{\(\underline{q_{i,t}^{load}},\,\overline{q_{i,t}^{load}}\)}{Minimum and maximum reactive power load at node \(i\) and time \(t\)}
\nomenclature[D,01]{\(y_{ij}\)}{Binary decision variable that is 1 if line \(ij\) is constructed and 0 otherwise}
\nomenclature[D,02]{\(y_{k}\)}{Binary decision variable that is 1 if RCS is constructed and 0 otherwise}
\nomenclature[D,03]{\(\beta_{ij}\)}{Binary decision variable that is 1 if node \(j\) is the parent of node \(i\) in the radial network}
\nomenclature[D,04]{\(F_{ij}\)}{Real-valued variable representing fictitious flow in line \(ij\)}
\nomenclature[D,05]{\(m_{k,u}^{se}\)}{Binary decision variable that is 1 if EV \(u\) is selected to charge at hub \(k\) and 0 otherwise}
\nomenclature[D,06]{\(p_{u,t}\)}{Charging power of EV \(u\) at time \(t\)}
\nomenclature[D,07]{\(n_{k}\)}{Number of EVs at hub \(k\)}
\nomenclature[D,08]{\(p_{k,u,t}^{ev}\)}{Charging power of EV $u$ at hub $k$ during time interval $t$}
\nomenclature[D,09]{\(p_{k,t}^{pv}\)}{Active power output of the PV unit at hub \(k\) in time \(t\)}
\nomenclature[D,10]{\(v_{i,t}\)}{Voltage magnitude at node \(i\) in time \(t\)}
\nomenclature[D,11]{\(e_{k,t}^{ess}\)}{Stored energy in the ESS at hub \(k\) in time \(t\)}
\nomenclature[D,12]{\(p_{ij,t},\,q_{ij,t}\)}{Active and reactive power flow in line \(ij\) at time \(t\)}
\nomenclature[D,13]{\(p_{k,t}^{ch},\,p_{k,t}^{dis}\)}{Charging and discharging power of the ESS at hub \(k\) in time \(t\)}
\nomenclature[D,14]{\(p_{t}^{sub},\,q_{t}^{sub}\)}{Active and reactive power output of the substation in time \(t\)}
\nomenclature[D,15]{\(p_{i,t}^{load},\,q_{i,t}^{load}\)}{Active and reactive power load at node \(i\) and time \(t\)}
\printnomenclature[1.45cm]
\vspace{-1.5em}
\section{Introduction}
\subsection{Background and Literature Review}
\IEEEPARstart{T}{he} electrification of transportation and the growing integration of renewable energy resources are fundamentally reshaping distribution networks (DN). The large-scale adoption of electric vehicles (EVs) introduces highly variable and stochastic charging demand, which can exacerbate peak loads, voltage fluctuations, and network congestion. At the same time, distributed photovoltaic (PV) generation and energy storage systems (ESS) provide opportunities to enhance local energy self-sufficiency and promote green mobility \cite{xiong2025two}. Renewable charging stations (RCS), which combine EV charging with PV and ESS, have therefore emerged as a promising solution to improve charging flexibility, facilitate renewable energy utilization, and reduce dependence on conventional DN resources. However, uncoordinated deployment of RCS can impose significant operational stress and necessitate costly infrastructure reinforcements. To address these challenges, coordinated planning of DN and RCS is essential, which enables efficient power dispatch, better load management, and cost-effective investment, while supporting the transition toward sustainable and resilient power systems \cite{lee2014electric}.

From modeling perspective, three main optimization frameworks have been widely applied to DN and EV charging station planning: stochastic optimization (SO), robust optimization (RO), and distributionally robust optimization (DRO). SO methods capture spatio-temporal uncertainties in charging demand and renewable generation through multi-scenario modeling, as demonstrated by Wang et al.\cite{wang2017stochastic}, Fu et al.\cite{faridimehr2018stochastic}, and Zare et al.\cite{zare2023stochastic}. RO methods focus on ensuring system stability under worst-case conditions, such as load fluctuations, uncertain EV charging demand, or possible DN disconnections, as studied by Arias et al.\cite{arias2017robust}, Zhao et al.\cite{zhao2023robust}, and Barhagh et al.\cite{barhagh2023optimal}. DRO seeks to combine the strengths of both approaches by enabling optimization without complete knowledge of probability distributions. This allows for tractable handling of spatio-temporal uncertainty in EV charging and renewable outputs, as illustrated in the works of Xiang et al.\cite{xiang2023distributionally}, Nguyen et al.\cite{nguyen2022decentralized}, and Shi et al.\cite{shi2022day}. Despite their contributions, most of these studies neglect the explicit modeling of RCS and fail to incorporate transportation flows and population distributions. Furthermore, RO methods are often overly conservative, SO methods may not ensure reliability under extreme conditions, and DRO approaches typically require restrictive confidence levels or assumptions about uncertainty distributions.

To bridge these gaps, Liu et al.\cite{liu2015stochastic} introduced the stochastic-robust optimization (SRO) framework, which balances the advantages of SO and RO while avoiding the restrictive assumptions required in DRO. SRO has since been applied to resilience management and unit commitment \cite{tsao2021sustainable, velloso2019two}, as well as disaster relief logistics \cite{bozorgi2013multi}. An important advancement is the recognition of \emph{decision-dependent uncertainty} (DDU), as introduced by Goel et al.\cite{goel2005lagrangean}, which distinguishes uncertainty influenced by planning decisions from \emph{decision-independent uncertainty} (DIU), which remains unaffected \cite{dupacova2006optimization}. In DN planning, DDU can be used to capture stochastic variations such as EV fleet sizes and charging patterns, while DIU accounts for robust operational uncertainties, such as demand fluctuations or renewable output variability. Few studies, however, have explored planning frameworks that integrate both DDU and DIU within an SRO setting.

On the algorithmic side, several decomposition and iterative refinement techniques have been developed. Benders decomposition \cite{bnnobrs1962partitioning} separates large-scale mixed-integer problems into a master problem and subproblems, facilitating tractable solutions. Zeng et al.\cite{zeng2013solving} introduced the column-and-constraint generation (C\&CG) method, which iteratively adds variables and constraints to improve solving efficiency. Zhao et al.\cite{zhao2012exact} extended this approach to binary-variable subproblems through the nested C\&CG (NC\&CG) algorithm. Tsang et al.\cite{tsang2023inexact} proposed the inexact C\&CG (iC\&CG) algorithm, which balances exploration and exploitation by allowing inexact subproblem solutions, thereby accelerating convergence while maintaining exactness in the overall solution. These algorithms have proven effective in solving RO and DRO models. For SRO problems, Sun et al.\cite{10659235} developed the parametric C\&CG (PC\&CG) algorithm, which improves convergence for multi-scenario problems under robust conditions. However, none of these methods are directly suited for solving tri-level SRO problems that explicitly incorporate both DDU and DIU. Conventional approaches often converge too slowly when applied to large-scale problems of this type. So we need new algorithm to address this challenge.
\vspace{-1.0em}
\subsection{Contributions and Paper Structure}
\begin{figure}[htbp]
    \centering
    \includegraphics[width=0.48\textwidth]{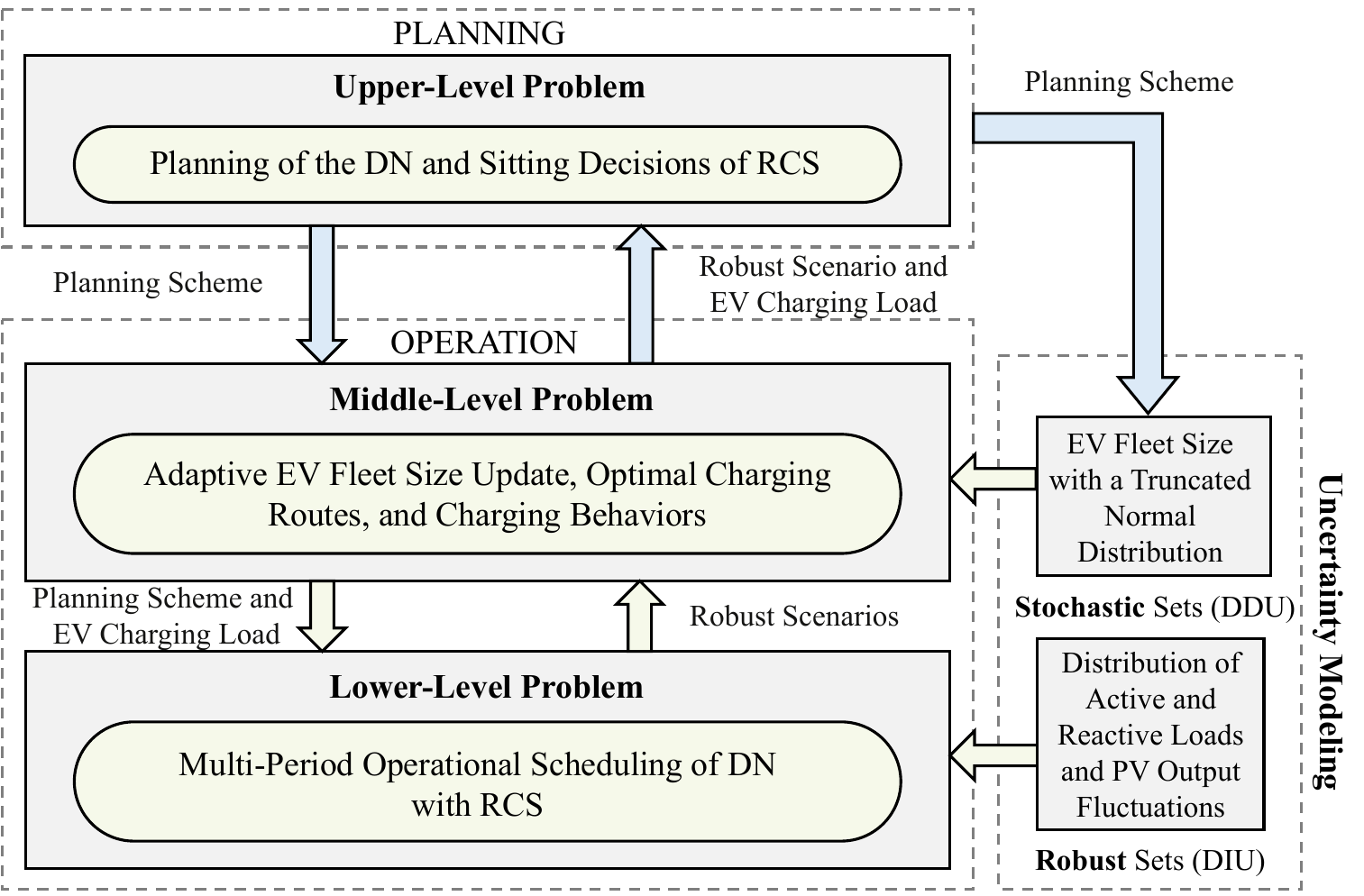}
    \caption{Framework of the tri-level stochastic-robust planning problem.}
    \label{fig1}
\end{figure}

To bridge the above gaps, this paper proposes a tri-level two-stage SRO framework for the integrated planning of RCS and DN, explicitly incorporating transportation flows and population distributions, as shown in Fig.~\ref{fig1}. The model distinguishes two types of uncertainties:  
(i) DDU refers to the EV charging load, which is influenced by RCS siting decisions. Additionally, the EV fleet size and initial position, both of which are stochastic, affect the charging load by determining the charging routes and behaviors. As a result, the EV charging load is also random, and DDU is represented as stochastic sets.
(ii) DIU encompasses load fluctuations and renewable generation variability, modeled as robust uncertainty sets that are independent of any planning decisions, including RCS siting or DN topology.
\vspace{-0.5em}
\vspace{-0.5em}

In long-term planning, EV fleet sizes are typically forecast within upper and lower bounds. Within this range, small deviations primarily affect queue lengths at RCS but do not significantly alter overall planning results. Evaluating fleet sizes within these bounds ensures that solutions remain practical, while queueing delays stay within acceptable limits for EV users. The main contributions of this paper are as follows:

1) A comprehensive model for joint planning of RCS and DN under multiple uncertainties, including EV fleet sizes, demand variations, and renewable generation fluctuations. The model explicitly incorporates population density and spatio-temporal traffic patterns to capture realistic EV charging behaviors.

2) A novel tri-level two-stage SRO framework is proposed, where DDU is represented in the middle-level problem as stochastic uncertainty sets, while DIU is addressed in the lower-level problem using robust optimization to safeguard operations against worst-case scenarios.

3) An adaptive inexact column-and-constraint generation (A-iC\&CG) algorithm is developed to solve the proposed model efficiently. Its convergence is proven to be controllable, with guarantees of finite-step convergence.

4) Numerical studies on a real-world case (Shenzhen, China) demonstrate that RCS siting correlates strongly with population density and substation proximity. The results also show that PV-EV charging stations represent the most cost-effective configuration, validating the efficiency and effectiveness of the proposed A-iC\&CG algorithm.

The remainder of this paper is organized as follows. Section~II presents the detailed formulation of the tri-level SRO model. Section~III describes the A-iC\&CG algorithm and its theoretical convergence properties. Section~IV illustrates the proposed approach through numerical case studies. Finally, Section~V concludes with key findings and directions for future research.
\vspace{-0.8em}
\subsection{Notations}
Let the sets $\mathbb{R}_{+}^{n}$ and $\mathbb{Z}_{+}^{n}$ represent the $n$-dimensional nonnegative real and integer vectors, respectively. Denote $k^{\underline{m}}$ as the falling factorial, defined as $k^{\underline{m}} = k(k-1)\cdots(k-m+1)$ for integers $k \ge m \ge 0$. The trace of a square matrix $A$ is written as $\operatorname{tr}(A)$. The vector $\mathbf{y} \in \{0,1\}^{n_{y_1}} \times \{0,1\}^{n_{y_2}}$ consists of binary decision variables of dimensions $n_{y_1}$ and $n_{y_2}$, whereas $\mathbf{z} \in \{0,1\}^{n_{z_1}} \times \mathbb{R}_+^{n_{z_2}}$ contains both binary variables of dimension $n_{z_1}$ and continuous nonnegative real variables of dimension $n_{z_2}$. A stochastic variable $v \sim TN(\mu, \sigma^{2}; [\underline{v}, \overline{v}])$ follows a truncated normal distribution with mean $\mu$, variance $\sigma^2$, and support restricted to $[\underline{v}, \overline{v}]$. The operator ``$\circ$'' denotes the Hadamard product: for vectors $\mathbf{a},\mathbf{b}\in\mathbb{R}^m$, $(\mathbf{a}\circ\mathbf{b})_i := a_i b_i$ for all $i$.
\vspace{-1.2em}
\section{Problem Formulation}
This section presents the mathematical formulation of the proposed planning problem. The distrbution network consists of \(n\) nodes, yielding \(n(n-1)/2\) candidate line corridors where exactly \(n-1\) lines are finally built to to form a radial topology. Node \(1\) is the primary substation connected to the utility grid. On transportation side, the system contains \(k\) hubs as candidate locations for RCS. If \(m\) RCS are to be deployed, there exist \(k^{\underline{m}}\) possible siting combinations.
\vspace{-0.5em}
\subsection{Tri-Level Two-stage Stochastic-Robust Planning Model}
\begin{figure}[htbp]
    \centering
    \includegraphics[width=0.48\textwidth]{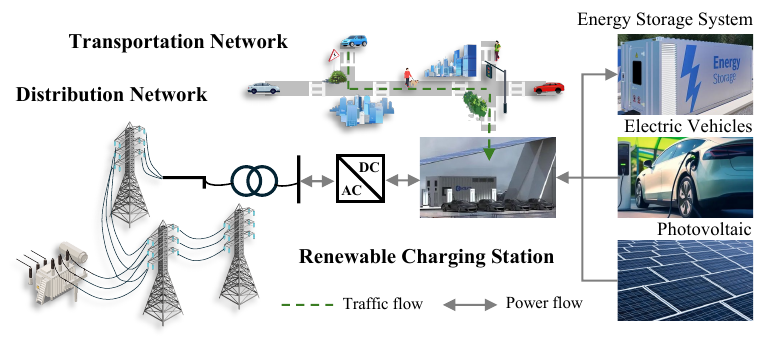}
    \vspace{-0.50em}
    \caption{Traffic-distribution coupled network with renewable charging stations.}
    \label{fig2}
\end{figure}

We develop a tri-level two-stage SRO model to minimize the overall costs of the transportation and distribution network, as shown in Fig.~\ref{fig2}. These costs encompass the construction and maintenance expenses for DN and RCS, as well as the cost of power losses. The model integrates both planning stage (upper-level problem) and operation stage (middle-level problem and lower-level problem).
\vspace{-1.0em}
\vspace{-0.5em}

\textit{1) Upper-Level Problem:}
The upper-level problem determines DN planning and RCS siting decisions, with the objective of minimizing infrastructure investment costs. The formulation is as follows:

\begin{gather}
\min_{y_{ij},y_{k}} \frac{d(1+d)^{\mathcal{T}_{Infr}}}{(1+d)^{\mathcal{T}_{Infr}}-1} (\sum_{ij \in \mathbb{L}} \varphi_{ij}y_{ij}+\sum_{k \in \mathbb{H}} \varphi_{k} y_{k}),\label{1}
\\
Infr\in\{Line,RCS\},\notag
\end{gather}
where \(d\) is the inflation rate, and \(\mathcal{T}_{\text{Infr}}\) is the lifespan of the corresponding infrastructure. Parameters \(\varphi_{ij}\) and \(\varphi_{k}\) denote the unit construction and maintenance costs of line \(ij\) and RCS \(k\), respectively, while \(y_{ij}\) and \(y_{k}\) are binary variables indicating construction decisions. The constraints are:
\begin{align}
& \beta_{ij}+\beta_{ji} = y_{ij},  
&& \forall (i,j) \in \mathbb{L} \label{father_son} \\
& \sum_{j \in \mathbb{N}/\{i\}} \beta_{ij} = 0,  
&& \forall i \in \mathbb{N}_{root} \label{father} \\
& \sum_{j \in \mathbb{N}/\{i\}} \beta_{ij} = 1,  
&& \forall i \in \mathbb{N}_{load} \label{son} \\
& \sum_{j \in \mathbb{N}/\{i\}} F_{ij}+D_i = \sum_{k \in \mathbb{N}/\{i\}} F_{ki},  
&& \forall i \in \mathbb{N}_{load} \label{Fflow} \\
& |F_{ij}| \leq y_{ij} M,  
&& \forall (i,j) \in \mathbb{L} \label{Fbulit}\\
& \sum_{k \in \mathbb{H}} y_{k} \geq m,  
&& \forall k \in \mathbb{H}. \label{RCS_min}
\end{align}

Constraints \eqref{father_son}--\eqref{son} enforce radiality through a spanning tree formulation. Binary variables \(\beta_{ij}\) define parent-child relationships: if \(\beta_{ij}=1\), node \(i\) is the parent of node \(j\); otherwise, either \(j\) is the parent of \(i\), or line \(ij\) is not constructed. Constraint \eqref{father} ensures that the substation node (\(\mathbb{N}_{\text{root}}=\{1\}\)) has no parent, while \eqref{son} guarantees that each load node has exactly one parent.
Constraints \eqref{Fflow}--\eqref{Fbulit} employ a single commodity flow formulation: fictitious flows \(F_{ij}\) ensure connectivity between all load nodes and the substation, preventing pseudo-loops. Here, \(D_i\) denotes the fictitious load at node \(i\), and \(M\) is a sufficiently large constant ensuring that \(F_{ij}=0\) whenever line \(ij\) is not constructed. The node set is defined as \(\mathbb{N}=\mathbb{N}_{root} \cup \mathbb{N}_{load}\), where $\mathbb{N}_{root}$ is the set of substation nodes and $\mathbb{N}_{load}$ is the set of load nodes. Finally, constraint \eqref{RCS_min} enforces that at least \(m\) RCS are constructed across the transportation hubs, ensuring sufficient charging capacity for EV demand.

\textit{2) Middle-Level Problem:}
Given the RCS siting decisions \(y_{k}\), the EV fleet sizes and initial positions follow a stochastic distribution. The fleet sizes are modeled by a truncated normal distribution within the forecasted bounds, while the initial positions are distributed according to population density. Each EV then selects an RCS for charging, which generates a decision-dependent charging load and forms the DDU for the overall problem. The middle-level objective aims to minimize the annualized routing and charging costs, formulated as:
\begin{gather}
\min_{m_{k,u}^{se},p_{u,t}} 365 c^{ev} \mathbf{\mathrm{\textit{tr}}}   \Bigl( (m_{k,u}^{in})^{T} R_{kk} m_{k,u}^{se} \Bigr) + 365 \sum_{\substack{t \in \mathbb{C}}} \sum_{\substack{u \in \mathbb{E}}} c_{t}^{tou} p_{u,t},\label{9}
\end{gather}
where $c^{ev}$ is the travel cost factor, and the stochastic indicator \(m_{k,u}^{in}\) reflects population density, denoting that EV \(u\) arrives at hub \(k\) requesting a charge. The matrix \(R_{kk}\) encodes inter-hub distances. The binary variable \(m_{k,u}^{se}\) equals 1 if EV \(u\) chooses to charge at hub \(k\), and 0 otherwise. The set \(\mathbb{C}\) collects charging intervals, and \(\mathbb{E}\) is the set of EVs. The parameter \(c_{t}^{tou}\) denotes the time-of-use (TOU) price at time $t$, and \(p_{u,t}\) is the charging power of EV \(u\) at time \(t\).
\begin{figure}[htbp]
    \centering
    \includegraphics[width=0.48\textwidth]{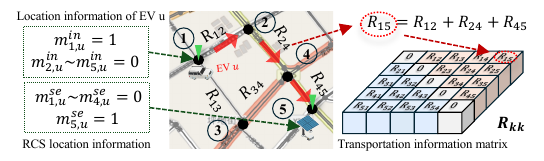}
    \vspace{-0.5em}
    \caption{Transportation network modeling illustration.}
    \label{fig3}
\end{figure}

For example, as shown in Fig.~\ref{fig3}, consider a transportation network with five hubs. Suppose EV \(u\) arrives at hub~1 when it needs to charge, so that \(m_{1,u}^{in} = 1\) and \(m_{k,u}^{in} = 0\) for \(k=2,\dots,5\). If a navigation application recommends the nearest RCS at hub~5 and EV \(u\) accepts, then \(m_{5,u}^{se}=1\) and \(m_{k,u}^{se}=0\) for \(k=1,\dots,4\). In the distance matrix \(R_{kk}\), the shortest path from hub~1 to hub~5 may be given by $R_{1,5} = R_{1,2} + R_{2,4} + R_{4,5}$, where \(R_{1,2}, R_{2,4},\) and \(R_{4,5}\) are entries of \(R_{kk}\). Summing the diagonal entries of \((m_{k,u}^{in})^{T} R_{kk} m_{k,u}^{se}\) then yields the total travel cost for all EVs in the network \cite{wang5185181integrated}.

EV fleet sizes are modeled by a truncated normal distribution. The initial spatial distribution of EVs and their charging parameters, e.g., state-of-charge (SoC) on arrival \(e_{u,o}\) and departure \(e_{u}\), are generated via Monte Carlo simulation, reflecting local population density, seasonal variation, and weekday/weekend distinctions \cite{wang5185181integrated}. This process yields realistic spatio-temporal traffic patterns. The constraints are:
\begin{align}
& \sum_{u\in\mathbb{E}} m^{se}_{k,u} = n_{k},
&& \forall k \in \mathbb{H} \label{9} \\
& \underline{n_{k}} \le n_{k} \le \overline{n_{k}},
&& \forall k \in \mathbb{H} \label{10} \\
& \sum_{k\in\mathbb{H}} m^{se}_{k,u} = 1,
&& \forall k \in \mathbb{H} \label{11} \\
& y_{k} \ge m^{se}_{k,u},
&& \forall k \in \mathbb{H} \label{12} \\
& 0 \le p^{ev}_{k,u,t}\le m^{se}_{k,u}\, M,
&& \forall k \in \mathbb{H},\, \forall t \in \mathbb{C} \label{13} \\
& -\,(1-m^{se}_{k,u})\, M \le p^{ev}_{k,u,t}-p_{u,t}\notag \\
& \le (1-m^{se}_{k,u})\, M,
&& \forall k \in \mathbb{H},\, \forall t \in \mathbb{C} \label{14} \\
& \underline{e_{u}} \le e_{u,0} + \sum_{t\in\mathbb{C}} p_{u,t} \le \overline{e_{u}},
&& \forall t \in \mathbb{C} \label{15} \\
& e_{u,0} + \sum_{t \in \mathbb{C}} p_{u,t} \ge e_{u},
&& \forall t \in \mathbb{C} \label{16} \\
& \sum_{t \in \mathbb{T}\setminus \mathbb{C}} \sum_{u \in \mathbb{E}} p_{u,t} = 0, 
&& \forall t \in \mathbb{C} \label{17} \\[3pt]
& \underline{p_{u}} \le p_{u,t} \le \overline{p_{u}},
&& \forall t \in \mathbb{C}. \label{18}
\end{align}

For all $u \in \mathbb{E}$, \(\mathbb{H}\) denotes the set of transportation hubs, \(\mathbb{T}\) the set of discrete time intervals, and \(\mathbb{C} \subseteq \mathbb{T}\) represents the charging periods. Constraints \eqref{9}--\eqref{10} ensure that the number of EVs assigned to hub \(k\) equals \(n_{k}\) and lies within feasible bounds. Constraints \eqref{11}--\eqref{12} require that each EV charges at exactly one hub, and only if an RCS is constructed there (\(y_{k}=1\)). 
Constraints~\eqref{13}--\eqref{14} introduce a big-$M$ linearization that links the binary assignment $m^{se}_{k,u}$ and the charging decision $p_{u,t}$ through the auxiliary variable $p^{ev}_{k,u,t}$. Specifically, if $m^{se}_{k,u}=1$, constraint~\eqref{14} forces $p^{ev}_{k,u,t}=p_{u,t}$; otherwise, constraint~\eqref{13} ensures $p^{ev}_{k,u,t}=0$. Constraints \eqref{15}--\eqref{16} bound the SoC of EV \(u\) between its minimum and maximum limits, ensuring safe operation and sufficient departure energy. Constraint \eqref{17} prohibits EVs from drawing power outside designated charging windows, while \eqref{18} confines charging power to allowable limits.

\textit{3) Lower-Level Problem:}
Given the planning decisions \(y_{ij}\), \(y_{k}\), and the EV charging load \(p_{k,t}^{ev}\), the lower-level problem minimizes the annualized power loss cost of the DN under the worst-case realization of DIU. The formulation is:
\begin{gather}
\max_{p_{i,t}^{load},\,q_{i,t}^{load},\,p_{k,t}^{pv}}
\min_{p_{ij,t},\, q_{ij,t}} 
365 \sum_{t \in \mathbb{T}} \sum_{ij \in \mathbb{L}} 
c_{t}^{tou} \, r_{ij} \frac{p_{ij,t}^2 + q_{ij,t}^2}{v_{i,t}^2}.
\label{19}
\end{gather}

In \eqref{19}, \(p_{i,t}^{load}\) and \(q_{i,t}^{load}\) represent the active and reactive loads at node \(i\) and time \(t\), while \(p_{k,t}^{pv}\) denotes the PV output at hub \(k\) and time \(t\). Their variability forms the DIU: $ \Bigl\{\,p_{i,t}^{load}, q_{i,t}^{load}, p_{k,t}^{pv} \,\Big|\,
\underline{p_{i,t}^{load}} \leq p_{i,t}^{load} \leq \overline{p_{i,t}^{load}}, \underline{q_{i,t}^{load}} \leq q_{i,t}^{load} \leq \overline{q_{i,t}^{load}}, \underline{p_{k,t}^{pv}} \leq p_{k,t}^{pv} \leq \overline{p_{k,t}^{pv}}\Bigr\}$,
where \(r_{ij}\) is the resistance of line \(ij\); \(p_{ij,t}\) and \(q_{ij,t}\) denote the active and reactive power flows on line \(ij\) at time \(t\); and \(v_{i,t}\) is the voltage magnitude at node \(i\). Since DN voltage typically remains close to 1 p.u. with only minor fluctuations, \(v_{i,t}\) is treated as a constant in the objective function for computational efficiency \cite{wang2023joint}.
\begin{align}
& e^{ess}_{k,t+1} = e^{ess}_{k,t} + p^{ch}_{k,t}\,\eta^{ess}_{ch,k} \notag \\
& - p^{dis}_{k,t}\,\eta^{ess}_{dis,k},
&& \forall k \in \mathbb{H}\ \label{20} \\
& \underline{p^{ch}_{k}} \le p^{ch}_{k,t} \le \overline{p^{ch}_{k}}, 
&& \forall k \in \mathbb{H}\ \label{21} \\
& \underline{p^{dis}_{k}} \le p^{dis}_{k,t} \le \overline{p^{dis}_{k}}, 
&& \forall k \in \mathbb{H} \label{22} \\
& \underline{e^{ess}_{k}} \le e^{ess}_{k,t} \le \overline{e^{ess}_{k}}, 
&& \forall k \in \mathbb{H}\ \label{23} \\
& p^{ch}_{k,t} \cdot p^{dis}_{k,t} = 0 \label{24}, 
&& \forall k \in \mathbb{H}\\
& p_{\sim i,t} = p_{i\sim,t} + p^{load}_{i,t} + \sum_{k \in \mathbb{H}(i)} y_k\notag \\
& \Bigl( \sum_{u\in\mathbb{E}} p^{ev}_{k,u,t} + p^{ch}_{k,t} - p^{dis}_{k,t} - p^{pv}_{k,t} \Bigr),
&& \forall i \in \mathbb{N}, k \in \mathbb{H} \label{25} \\
& q_{\sim i,t} = q_{i\sim,t} + q^{load}_{i,t}, 
&& \forall i \in \mathbb{N} \label{26} \\
& |v_{i,t}^{2} - v_{j,t}^{2} - 2(r_{ij}p_{ij,t} + x_{ij}q_{ij,t})| \notag \\
& \le (1-y_{ij})M, 
&& \forall ij \in \mathbb{L} \label{27} \\
& p_{ij,t}^2 + q_{ij,t}^2 \le y_{ij}\, \overline{s_{ij}}^2, 
&& \forall ij \in \mathbb{L} \label{28} \\
& \underline{v} \le v_{i,t} \le \overline{v}, 
&& \forall i \in \mathbb{N} \label{29} \\
& \underline{p^{sub}} \le p^{sub}_{t} \le \overline{p^{sub}}, 
&&  \label{30} \\
& \underline{q^{sub}} \le q^{sub}_{t} \le \overline{q^{sub}}.
&& \label{31}
\end{align}

For all $t \in \mathbb{T}$, constraint \eqref{20} ensures an energy balance for the ESS, where the energy level at each time step is updated based on the previous energy state and any charging or discharging actions, with \(\eta_{ch,k}^{ess}\) and \(\eta_{dis,k}^{ess}\) representing the charging and discharging efficiencies, respectively. Constraints \eqref{21}--\eqref{22} limit the charging power \(p_{k,t}^{ch}\) and discharging power \(p_{k,t}^{dis}\) within specified bounds, which are defined by the charging and discharging limits \(\overline{p_{k}^{ch}}\), \(\underline{p_{k}^{ch}}\), \(\overline{p_{k}^{dis}}\), and \(\underline{p_{k}^{dis}}\). Furthermore, constraint \eqref{23} ensures that the energy storage at each hub remains within its predefined limits, \(\overline{e_{k}^{ess}}\) and \(\underline{e_{k}^{ess}}\), while constraint \eqref{24} prohibits simultaneous charging and discharging actions \cite{li2015sufficient}. Besides, constraints \eqref{25}--\eqref{26} represent the active and reactive power balances at each node, where \(\mathbb{H}(i) := \{ k \in \mathbb{H} \mid \text{hub $k$ is connected to DN node $i$}\}\). Variables \(p_{\sim i,t}\) and \(p_{i\sim,t}\) denote active power entering and leaving node \(i\), while \(q_{\sim i,t}\) and \(q_{i\sim,t}\) denote the corresponding reactive power. Constraint \eqref{27} enforces the voltage drop equation between two connected nodes, where the line reactance is denoted by \(x_{ij}\). Constraint \eqref{28} enforces the line capacity limit, and equation \eqref{29} restricts the voltage fluctuations to acceptable limits. Constraints \eqref{30}--\eqref{31} specify the active and reactive power bounds at the substation, where \(\overline{s_{ij}}\) is the maximum line capacity, and \(p_{t}^{sub}\), \(q_{t}^{sub}\) are the active and reactive powers at the substation.

For ESS in RCS, two conditions are satisfied: (i) No-arbitrage on marginal prices, and (ii) Charging is cheaper than the node price. As a result, constraint \eqref{24} can be relaxed, and it has been proven that this relaxation is exact \cite{li2015sufficient}.
\vspace{-1.5em}
\subsection{Compact Formulation and Model Reformulation}
For the sake of notational clarity and analytical tractability, the proposed tri-level optimization problem is expressed in the following compact form:
\begin{gather}
\min_{\mathbf{y} \in \mathbb{Y}} \mathbf{c^{\mathsf T}y} + \min_{\substack{\mathbf{v}_l \in \mathbb{V} \\ \mathbf{z}_l \in \mathcal{F}(\mathbf{y}, \mathbf{v}_l)}} \mathbf{d}_l^{\mathsf T} \mathbf{z}_l +
\max_{\mathbf{u}_s \in \mathbb{U}} \min_{\mathbf{x}_{s,l} \in \mathcal{G}(\mathbf{y}, \mathbf{z}_l, \mathbf{u}_s)}\frac{1}{2}\,\mathbf{x}_{s,l}^{\mathsf T}\mathbf{Q}_{s,l}\,\mathbf{x}_{s,l}, \label{32}
\end{gather}
\vspace{-1.0em}
where for each stochastic scenario \(l\) and robust scenario \(s\):
\begin{align*}
\mathbb{Y} &= \left\{ \mathbf{y} \in \{0,1\}^{n_{y_1}} \times \{0,1\}^{n_{y_2}} \mid \mathbf{A y} \le \mathbf{b} \right\}, \\
\mathbb{V} &= \left\{ \mathbf{v}_l \in \mathbb{Z}_{+}^{n_{v}} \mid \mathbf{v}_l \sim TN\bigl(\mu,\sigma^2;[\underline{v},\overline{v}]\bigl) \right\}, \\
\mathcal{F} &=\left\{\mathbf{z}_l \in \{0,1\}^{n_{z_1}} \times \mathbb{R}_{+}^{n_{z_2}} \mid \mathbf{H}_l \mathbf{z}_l + \mathbf{h}_l \le \mathbf{L}_l \mathbf{v}_l + \mathbf{F}_l \mathbf{y} \right\},\\
\mathbb{U} &= \left\{ \mathbf{u}_s \in \mathbb{R}_{+}^{n_{u}} \mid \mathbf{B}_s \mathbf{u}_s \le \mathbf{e}_s \right\}, \\
\mathcal{G} &= \left\{\mathbf{x}_{s,l} \in \mathbb{R}_{+}^{n_{x}} \mid \mathbf{I}_{s,l} \textbf{x}_{s,l} + \mathbf{j}_{s,l} \le \mathbf{M}_{s,l} \mathbf{u}_{s} + \mathbf{N}_{s,l} \mathbf{z}_{l} + \mathbf{K}_{s,l} \mathbf{y} \right\},
\end{align*}
where \(\mathbf{y}\) represents the upper-level planning vector, including line construction variables \(y_{ij}\), RCS siting variables \(y_{k}\), and auxiliary binary variables \(\beta_{ij}, F_{ij}\). The stochastic variable \(\mathbf{v}_l\) represents EV fleet sizes, sampled from a truncated normal to generate scenarios; \(\mathbf{z}_l\) is the middle-level decision vector, including \(m_{k,u}^{se}, p_{k,t}^{ev}, p_{u,t}\). The robust variable \(\mathbf{u}_s\) corresponds to PV output \(p_{k,t}^{pv}\) and active/reactive power \(p_{i,t}^{load}, q_{i,t}^{load}\). The lower-level operational vector \(\mathbf{x}_{s,l}\) includes \(e_{i,t}^{ess}, p_{i,t}^{ch}, p_{i,t}^{dis}, p_{i,t}^{pv}, p_{ij,t}, q_{ij,t}, v_{i,t}, p_{t}^{sub}, q_{t}^{sub}\). The dimensions \(n_{y_1}, n_{y_2}, n_{v}, n_{z_1}, n_{z_2}, n_{u}, n_{x}\) are appropriate quantities standing for dimensions of $\mathbf{y}, \mathbf{v}_l, \mathbf{z}_l, \mathbf{u}_s, \mathbf{x}_{s,l}$, respectively. Positive semidefinite weighting $\mathbf{Q}_{s,l}\succeq \mathbf{0}$, coefficient vectors $\mathbf{c}, \mathbf{d}_l, \mathbf{b}, \mathbf{h}_l, \mathbf{e}_s, \mathbf{j}_{s,l}$, and constraint matrices $\mathbf{A}, \mathbf{H}_l, \mathbf{L}_l, \mathbf{F}_l, \mathbf{B}_s, \mathbf{I}_{s,l}, \mathbf{M}_{s,l}, \mathbf{N}_{s,l}$ and $\mathbf{K}_{s,l}$ are all with compatible dimensions. Note that all vectors are column vectors and appear in boldface, matrices are bold uppercase.

For fixed $(\mathbf{y},\mathbf{z}_l,\mathbf{u}_{s})$ and a given pair $(s,l)$, the lower-level problem is quadratic programming problem: $\{ \min_{\mathbf{x}_{s,l} \in \mathbb{R}_{+}^{n_x}} \frac{1}{2}\,\mathbf{x}_{s,l}^{\mathsf T}\mathbf{Q}_{s,l}\,\mathbf{x}_{s,l} \mid \mathbf{I}_{s,l} \textbf{x}_{s,l} + \mathbf{j}_{s,l} \le \mathbf{M}_{s,l} \mathbf{u}_{s} + \mathbf{N}_{s,l} \mathbf{z}_{s,l} + \mathbf{K}_{s,l} \mathbf{y}\}$. Since Slater’s condition holds and $\mathbf{Q}_{s,l}\succeq\mathbf{0}$, the KKT conditions are necessary and sufficient. 
Let $\boldsymbol{\pi}_{s,l}$ be multipliers for the linear inequalities and 
$\boldsymbol{\lambda}_{s,l}$ for $\mathbf{x}_{s,l}\ge\mathbf{0}$. 
The Lagrangian is $\mathcal{L}(\mathbf{x}_{s,l},\boldsymbol{\pi}_{s,l},\boldsymbol{\lambda}_{s,l})
=\frac{1}{2}\,\mathbf{x}_{s,l}^{\mathsf T}\mathbf{Q}_{s,l}\,\mathbf{x}_{s,l}
+\boldsymbol{\pi}_{s,l}^{\mathsf T}\left(\mathbf{I}_{s,l}\mathbf{x}_{s,l}+\mathbf{j}_{s,l}
-\mathbf{M}_{s,l}\mathbf{u}_s-\mathbf{N}_{s,l}\mathbf{z}_l-\mathbf{K}_{s,l}\mathbf{y}\right)- \boldsymbol{\lambda}_{s,l}^{\mathsf T}\mathbf{x}_{s,l}.$ Combining the outer maximization over $\mathbf{u}_s$ with these KKT conditions yields the equivalent single-level maximization:
\begin{gather}
\max_{\mathbf{u}_s,\ \mathbf{x}_{s,l},\ \boldsymbol{\pi}_{s,l}, \boldsymbol{\lambda}_{s,l},\ \mathbf{o}_{s,l},\ \mathbf{w}_{s,l}}\frac{1}{2}\,\mathbf{x}_{s,l}^{\mathsf T}\mathbf{Q}_{s,l}\,\mathbf{x}_{s,l}
\label{33}
\end{gather}
\begin{align}
& \mathbf{B}_s\,\mathbf{u}_s \le \mathbf{e}_s, && \label{34} \\
& \mathbf{I}_{s,l}\mathbf{x}_{s,l} + \mathbf{j}_{s,l} \le \mathbf{M}_{s,l}\mathbf{u}_s + \mathbf{N}_{s,l}\mathbf{z}_l + \mathbf{K}_{s,l}\mathbf{y}, && \label{35} \\
&  \mathbf{Q}_{s,l}\mathbf{x}_{s,l} + \mathbf{I}_{s,l}^{\mathsf T}\boldsymbol{\pi}_{s,l} - \boldsymbol{\lambda}_{s,l} = \mathbf{0}, && \label{36} \\
& \boldsymbol{\pi}_{s,l} \ge \mathbf{0}, \quad \boldsymbol{\lambda}_{s,l} \ge \mathbf{0}, && \label{37} \\
& \mathbf{I}_{s,l}\mathbf{x}_{s,l} + \mathbf{j}_{s,l} - \mathbf{M}_{s,l}\mathbf{u}_s - \mathbf{N}_{s,l}\mathbf{z}_l - \mathbf{K}_{s,l}\mathbf{y} \notag \\
& \ge - M(\mathbf{1} - \mathbf{o}_{s,l}), && \label{38} \\
& \mathbf{0} \le \boldsymbol{\pi}_{s,l} \le M \mathbf{o}_{s,l}, && \label{39} \\
& \mathbf{0} \le \mathbf{x}_{s,l} \le M \mathbf{w}_{s,l}, && \label{40} \\
& \mathbf{0} \le \boldsymbol{\lambda}_{s,l} \le M (\mathbf{1} - \mathbf{w}_{s,l}), && \label{41} \\
& \mathbf{o}_{s,l} \in \{0,1\}, \quad \mathbf{w}_{s,l} \in \{0,1\}, && \label{42}
\end{align}
where equation \eqref{35} represents primal feasibility for the optimization problem, ensuring that the constraints related to the decision variables are satisfied. Equation \eqref{36} represents stationarity, ensuring the optimality condition for the lagrangian with respect to the primal variables \(\mathbf{x}_{s,l}\) and the dual variables \(\boldsymbol{\pi}_{s,l}\) and \(\boldsymbol{\lambda}_{s,l}\). 
Equation \eqref{37} represents dual feasibility, which ensures that the dual variables \(\boldsymbol{\pi}_{s,l}\) and \(\boldsymbol{\lambda}_{s,l}\) remain non-negative.
Constraints \eqref{38}--\eqref{39} enforce the complementarity conditions, ensuring the primal and dual variables are complementary. Specifically, these constraints enforce the condition:
\[
\boldsymbol{\pi}_{s,l} \circ \left( \mathbf{I}_{s,l} \mathbf{x}_{s,l} + \mathbf{j}_{s,l} - \mathbf{M}_{s,l} \mathbf{u}_s - \mathbf{N}_{s,l} \mathbf{z}_l - \mathbf{K}_{s,l} \mathbf{y} \right) = \mathbf{0}.
\]
These complementarity conditions are enforced through the big-\(M\) method, using the binary variables \(\mathbf{o}_{s,l}\).
Constraints \eqref{40} and \eqref{41} introduce binary variables \(\mathbf{w}_{s,l}\) to linearize the complementarity condition \(\boldsymbol{\lambda}_{s,l} \circ \mathbf{x}_{s,l} = 0\). These binary variables \(\mathbf{w}_{s,l}\) ensure that \(\boldsymbol{\lambda}_{s,l}\) and \(\mathbf{x}_{s,l}\) are complementary.
\vspace{-1.0em}
\section{Adaptive Inexact Column-and-Constraint Generation Algorithm}
This section presents the proposed A-iC\&CG algorithm designed to solve the tri-level two-stage SRO problem that simultaneously accounts for DDU and DIU. The algorithm guarantees controllable convergence and ensures that the optimal solution to be found within a finite number of iterations.
\vspace{-1.5em}
\subsection{A-iC\&CG Algorithm for Solving Two-Stage SRO}
Table \ref{RD} summarizes the proposed A-iC\&CG algorithm. After solving the optimization problem within a prescribed optimality gap, a commercial solver returns both an upper and a lower bound. The upper bound corresponds to a feasible solution that satisfies all constraints, whereas the lower bound is obtained from a relaxed formulation in which some constraints are omitted. As a result, the upper bound is regarded as the practical solution of interest, while the lower bound mainly serves as a reference for evaluating convergence.

In Step~1, the algorithm initializes all parameters, where $\overline{UB}$ and $\overline{LB}$ denote the global upper and lower bounds of the problem, $i$ is the iteration index, and $k$ tracks the effective iteration count. In Step~2, the upper-level problem is solved within a tolerance $\varepsilon_{up}^i$, producing the bounds $(LB^i, UB^i)$ at the $i$-th iteration. If the solution is valid, i.e., $LB^i \geq \overline{LB}$, the effective iteration count $k$ is updated and $\overline{LB}$ is refreshed to accelerate convergence. Step~3 adaptively updates the truncated distribution parameters of $\mathbf{v}_l^i$ and solves the middle-level problem to obtain the decision-dependent EV charging load $\mathbf{z}_l^i$. Step~4 then solves the lower-level problem under DIU and updates the global upper bound $\overline{UB}$. Step~5 performs the optimality test by comparing $\overline{UB}$ and $UB^k$. If their gap is below the threshold $\varepsilon$, the algorithm terminates with solution $\mathbf{y}^i$. Otherwise, two paths are possible: in the exploitation phase, the tolerance is reduced to $\alpha \varepsilon_{up}^i$ to refine the solution; in the exploration phase, Step~6 enlarges the DIU set and the algorithm proceeds.
\vspace{-1.0em}
\begin{table}[htbp]
\vspace{-0.5em}
\caption{A-iC\&CG Algorithm for Solving Two-Stage SRO}
\vspace{-0.5em}
\label{RD}
\scriptsize
\begin{tabular}{ll}
\hline \hline
\textbf{Step 1} 
& \textbf{Initialization:} Set $\overline{LB}=0$, $\overline{UB}=+\infty$, $i=1$, $k=1$, $\mathbb{U}=\emptyset$. 
\\ &Choose parameters $\varepsilon \in [0,1]$, $\tilde{\varepsilon} \in (0,\varepsilon/(\varepsilon+1))$, 
$\varepsilon^{i}_{up}$, $\alpha \in (0,1)$.\\
\textbf{Step 2} 
& \textbf{Upper-Level Problem:} \\
& Within an optimality gap $\varepsilon_{up}^{i}$, solve:
\\
& $\begin{array}{l}
\qquad \qquad \qquad \qquad \min_{\mathbf{y} \in \mathbb{Y}} \quad \mathbf{c^{\mathsf T}y} + \eta,\\
  \text{s.t. } 
  \begin{cases}
  \eta \ge \max_{\mathbf{u}_s, \mathbf{x}_{s,l}, \boldsymbol{\pi}_{s,l}, \boldsymbol{\lambda}_{s,l}, \mathbf{o}_{s,l}, \mathbf{w}_{s,l}}
  \tfrac{1}{2}\mathbf{x}_{s,l}^{\mathsf T}\mathbf{Q}_{s,l}\mathbf{x}_{s,l},\\
  \mathbf{c}^{\mathsf T}\mathbf{y} + \eta \ge \overline{LB}, \\
  \mathbf{A}\mathbf{y} \le \mathbf{b}, \\
  \text{Constraints (34)--(42).}
\end{cases}
\end{array} $    \\  
& Record the best feasible solution $(\mathbf{y}^i, \eta^i)$.
Record $LB^i \ge \overline{LB}$ and \\
& set $UB^i = \mathbf{c}^{\mathsf T}\mathbf{y}^i + \eta^i$.
  If $LB^i \ge \overline{LB}$, set $k=i$. \\
& Update $\overline{LB} = UB^i$.
\\
\textbf{Step 3} 
& \textbf{Middle-Level Problem:} \\
& \textbf{3.1} Adaptive Oracle Tracking:\\
& If from Step 2, sample \( \mathbf{v}_l^i \sim TN(\mu^i, {\sigma^i}^2; [\underline{v}, \mu^{i-1}]) \), \\
& proceed to Step 3.2 then Step 4. \\
& If from Step 6, sample \( \mathbf{v}_l^i \sim TN(\mu^i, {\sigma^i}^2; [\mu^{i-1}, \overline{v}]) \), \\
& return to Step 2 after Step 3.2. \\
& \textbf{3.2} Given $(\mathbf{y}^i,\mathbf{v}_l^i)$, solve:
\\
& $\begin{array}{l}
\qquad \qquad \qquad \qquad \min_{\substack{\mathbf{z}_l \in \mathcal{F}(\mathbf{y}^i, \mathbf{v}^i_l)}} \mathbf{d}_l^{\mathsf T} \mathbf{z}_l,\\
\qquad \qquad \quad \text { s.t.} 
 \quad \mathbf{H}_l \mathbf{z}_l + \mathbf{h}_l \le \mathbf{L}_l \mathbf{v}^i_l + \mathbf{F}_l \mathbf{y}^i. \\
\end{array} $    \\  
& Record the optimal solution $\mathbf{z}_l^i$ and stochastic scenario $\mathbf{v}_l^i$.\\
\textbf{Step 4} 
& \textbf{Lower-Level Problem:} \\
& For fixed $(\mathbf{y}^i,\mathbf{z}_l^i)$, solve:
\\
& $\begin{array}{l}
\quad \max_{\mathbf{u}_s,\ \mathbf{x}_{s,l},\ \boldsymbol{\pi}_{s,l}, \boldsymbol{\lambda}_{s,l},\ \mathbf{o}_{s,l},\ \mathbf{w}_{s,l}}\frac{1}{2}\,\mathbf{x}_{s,l}^{\mathsf T}\mathbf{Q}_{s,l}\,\mathbf{x}_{s,l},\\
\qquad \qquad \qquad \text { s.t.} 
\quad \textnormal{Constraints (34)--(42)}. \\
\end{array} $    \\  
& Obtain the optimal $(\mathbf{x}_{s,l}^i,\mathbf{u}_s^i)$ with objective $D^i$.\\
& Update $\overline{UB} = \min\{\overline{UB}, \mathbf{c}^{\mathsf T}\mathbf{y}^i + D_i\}$.\\
\textbf{Step 5} 
& \textbf{Optimality Test and Backtracking:}  \\
& If $(\overline{UB}-LB^k)/\overline{UB}<\varepsilon$, then terminate and return $\mathbf{y}^i$; \\
& Otherwise, proceed as follows:\\
& \noindent $\bullet$ \textbf{Exploitation:} if $(\overline{UB} - UB^i)/\overline{UB} < \tilde{\varepsilon}$, set $i = k$, $\overline{LB} = LB^k$, \\
& and update $\varepsilon_{up}^{i} = \alpha \,\varepsilon_{up}^{i}$ for all $i \ge k$. Then go back to Step 2.\\
& \noindent $\bullet$ \textbf{Exploration:}  if $(\overline{UB}-UB^i)/\overline{UB}
\geq \tilde{\varepsilon}$, then go to Step 6.\\
\textbf{Step 6} 
& \textbf{Robust Scenario Set Enlargement:} \\
& Update $\mathbb{U} = \mathbb{U} \cup \{\,\mathbf{u}_{s}^i\}$.\\
& Set $i = i + 1$ and return to Step 3.\\
\hline \hline 
\end{tabular}
\end{table}
\vspace{-0.5em}

\begin{figure}[htbp]
    \centering
    \includegraphics[width=0.45\textwidth]{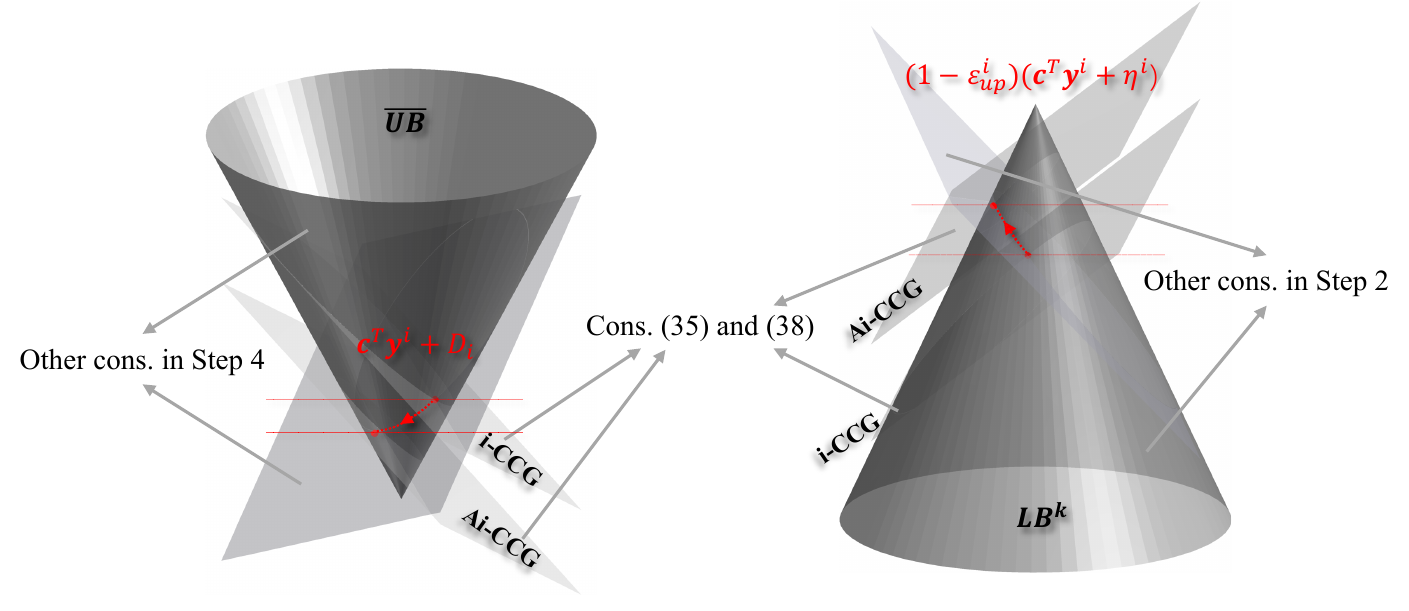}
    \caption{Variation of problem bounds based on upper and lower-level solutions.}
    \label{fig5}
\end{figure}

Compared with the C\&CG algorithm (Section 3 in reference \cite{zeng2013solving}) and the iC\&CG algorithm \cite{tsang2023inexact}, the proposed A-iC\&CG algorithm introduces \emph{Adaptive Oracle Tracking} to address DDU. When an iteration is effective, the global upper bound is updated as $\overline{UB} = \mathbf{c}^{\mathsf T}\mathbf{y}^i + D_i$ after solving the lower-level problem, while the effective lower bound is updated as $LB^k = (1-\varepsilon_{up}^{i})(\mathbf{c}^{\mathsf T}\mathbf{y}^i + \eta^i)$ after solving the upper-level problem. Since the objective function value is positively correlated with $\mathbf{z}_l^i$, the adjustment of $\mathbf{z}_l^i$ through the sampled $\mathbf{v}_l^i$ affects the bound updates. Specifically, because $\mathbf{z}_l^i$ appears in constraints~(35) and~(38), these constraints ensure that $\overline{UB}$ decreases and $LB^k$ increases, thereby tightening the feasible region and narrowing the bound gap. This adaptive mechanism reduces the number of iterations and conserves computational effort by dynamically tuning the distribution intervals of the random variables. Consequently, each iteration of A-iC\&CG produces a lower $\overline{UB}$ and a higher $LB^k$ compared with the iC\&CG algorithm. The variation of problem bounds is illustrated in Fig.~\ref{fig5}. For completeness, the iC\&CG algorithm for solving the two-stage RO problem is provided in Table~\ref{RO} in the Appendix.
\vspace{-1.0em}
\vspace{-0.5em}

\noindent \textbf{Remark 1.}
After solving the upper-level problem with a prescribed gap, the corresponding bounds of the two-stage SRO are obtained, namely the upper bound $UB^i$ and lower bound $LB^i$ from the $i$-th iteration. Once the lower-level problem is solved, the global upper bound $\overline{UB}$ of the two-stage SRO is updated. If the $i$-th iteration is effective, the effective iteration index is set as $k=i$. Together, $\overline{UB}$ and $LB^k$ provide a reliable measure of the degree of convergence of the two-stage SRO problem.
\vspace{-1.5em}
\subsection{Theoretical Properties of the A-iC\&CG Algorithm}
This section examines the theoretical properties of the A-iC\&CG algorithm, focusing on the convergence behavior and solution accuracy of the algorithm.

\noindent \textbf{Proposition 1.}
Assume the A-iCCG algorithm first enters an \emph{Exploration} phase and then switches to \emph{Exploitation}, while each upper-level problem retains an optimality gap \( \varepsilon_{up}^i \). The actual relative gap of two-stage SRO is bounded.

\noindent \textbf{Proof.} Motivated by reference \cite{tsang2023inexact}, we consider two scenarios.
(\textit{i})~If $i=k$, then from $\bigl(UB^k-LB^k\bigr)/UB^k\le\varepsilon_{up}^k$ it follows $UB^k(1-\varepsilon_{up}^k)\le LB^k$. Meanwhile, the exploitation condition $(\overline{UB}-UB^k)/\overline{UB}<\tilde{\varepsilon}$ gives $UB^k>(1-\tilde{\varepsilon})\,\overline{UB}$. Combining implies $\overline{UB}<(UB^k)/(1-\tilde{\varepsilon})\le LB^k/\bigl((1-\varepsilon_{up}^k)(1-\tilde{\varepsilon})\bigr)$, so $(\overline{UB}-LB^k)/\overline{UB}\le1-(1-\tilde{\varepsilon})(1-\varepsilon_{up}^k)$. 
(\textit{ii})~If $i>k$, then over multiple exploitations from $n=k$ to $i-1$, one has $\bigl(UB^{n+1}-LB^{n+1}\bigr)/UB^{n+1}\le\varepsilon_{up}^{n+1}$ implying $UB^{n+1}(1-\varepsilon_{up}^{n+1})\le LB^{n+1}=UB^n$  for same $\mathbf{z}_l$ and $\mathbb{U}$. Iterating yields $UB^i \le UB^k\,\prod_{n=k+1}^i (1-\varepsilon_{up}^n)^{-1}$. From $n=k$ similarly $UB^k \le LB^k/(1-\varepsilon_{up}^k)$, so $UB^i\le LB^k\prod_{n=k}^i(1-\varepsilon_{up}^n)^{-1}$. Meanwhile, an exploitation condition $(\overline{UB}-UB^i)/\overline{UB}<\tilde{\varepsilon}$ implies $UB^i>(1-\tilde{\varepsilon})\,\overline{UB}$, and gives $(\overline{UB}-LB^k)/\overline{UB}\le1 -(LB^k/UB^i)\,(UB^i/\overline{UB}) \le 1 -\bigl(\prod_{n=k}^i (1-\varepsilon_{up}^n)^{-1}\bigr)^{-1}(1-\tilde{\varepsilon})$. Combining both cases shows the gap remains bounded by the stated maximum, thus ensuring controllable convergence.

\noindent \textbf{Remark 2.}
Due to the stochastic-robust nature of the problem, the lower bound \( LB^i \) obtained from the upper-level problem may occasionally exceed the upper bound \( \overline{UB} \) by a small, acceptable margin. This may occur when random factors in the \emph{Adaptive Oracle Tracking} phase adjust the feasible domain. However, any sampled value within \( [\underline{v}, \overline{v}] \) remains valid for the stochastic variable, and the SRO is still considered convergent under these conditions.

\noindent \textbf{Proposition 2.}
The A-iC\&CG algorithm converges in finite time.

\noindent \textbf{Proof.}
Every \emph{Exploration} step expands $\mathbb{U}$, which is finite or a polytope with finitely many extreme points, so \emph{Exploration} can occur only finitely often.
Suppose, for contradiction, that the upper-level problem is solved infinitely often under some fixed $\mathbb{U}$ and $\mathbf{z}_l$. This would imply perpetual \emph{Exploitation} without termination or return to \emph{Exploration}, effectively keeping $i=k$ for every iteration. From Proposition~1, $(\overline{UB}-LB^k)/\overline{UB} \,\le\, 1 - \bigl(1-\tilde{\varepsilon}\bigr)\bigl(1-\varepsilon_{up}^k\bigr)$. As iterations continue, $\varepsilon_{up}^k \to 0$, which forces $(\overline{UB}-LB^k)/\overline{UB} \,\le\, \tilde{\varepsilon}$. Since $\tilde{\varepsilon} \,\le\, \varepsilon/(\varepsilon+1) \,\le\, \varepsilon$, this contradicts the termination criterion $(\overline{UB}-LB^k)/\overline{UB} < \varepsilon$. Therefore, no scenario subset $\mathbb{U}$ or \(\mathbf{z}_l\) can lead to infinitely many upper-level re-solves.
Combining these two arguments proves that Ai-CCG converges in finite time.

\noindent
\textbf{Remark 3.}
Given the structure of our tri-level SRO model, the proposed A-iC\&CG algorithm can solves problems where the middle-level problem involves stochastic DDUs, and the lower-level problem involves robust DIUs. Additionally, if the middle-level problem incorporates a robust uncertainty set while the lower-level problem contains a stochastic uncertainty set, the tri-level SRO problem can be solved using the PC\&CG algorithm (Section III in reference \cite{10659235}).
\vspace{-0.8em}
\section{Case Studies}
This section presents simulation results to validate the A-iC\&CG algorithm in solving the tri-level two-stage SRO problem. We compare the planning results and convergence speed with other algorithms, and also analyze the sensitivity of DDU and DIU.
\vspace{-1.0em}
\subsection{Case Description}
The proposed model is evaluated on a coupled 47-node DN and a 68-hub transportation system in Longgang District, Shenzhen, China, spanning residential, office, commercial, and industrial areas \cite{wang5185181integrated}. The construction and operational costs of the integrated system are detailed in Table~\ref{tab:construction costs}, which includes three types of RCS. The table also outlines the peak, off-peak, and standard (p/o/s) pricing categories for Shenzhen \cite{liu2022coordinated}. The RCS (PV-ESS-EV) includes 75~kW PV array and 1.5~MWh ESS, with maximum charging and discharging rates of 0.2~MW and 0.3~MW, respectively, and efficiency ratios of 0.9 and 1.1. Over the next 20 years, the number of EVs in the transportation system is forecasted to fluctuate between 2987 and 4011. PV installations in the area are subsidized at 1.02 CNY/W, and an electricity usage subsidy of 0.05 CNY/kWh is provided. Additionally, the inflation rate is set at 0.05, and voltage magnitude is constrained between 0.9 p.u. and 1.1 p.u. to maintain acceptable levels. The model was formulated using the YALMIP tool in MATLAB (2021a) and evaluated with the GUROBI Optimizer (9.5.2) on the Apple M3 Pro chip, which boasts a 12-core CPU and an 18-core GPU.
\begin{table}[htbp]
  \centering
  \vspace{-0.5em}
  \caption{Construction and Operation Costs ($10^{4}$ CNY \textyen)}
  \vspace{-1.0em}
  \label{tab:construction costs}
  \tabcolsep=5pt
  \begin{tabular}{ccccc}
    \hline \hline
    \thead{Title}          & \thead{Type}  & \thead{Candidate} & \thead{Cost} & \thead{Lifespan} \\ \hline
    \multirow{3}{*}{RCS}   & PV-ESS-EV & 1 -- 68 hubs & 344.50 & 20/30/25 \\
                           & PV-EV         & 1 -- 68 hubs & 194.50 & 20/30 \\
                           & EV-only       & 1 -- 68 hubs & 187.00 & 20 \\ \hline
    \multirow{2}{*}{EV}    & consumption   & RCS           & 115 Wh/km  & - \\
                           & frequency     & RCS           & 0.15 times/day & - \\ \hline
    \multirow{1}{*}{TOU}   & p/o/s & all day & \multicolumn{2}{c}{$1.11/0.65/0.25\times10^{-4}\text{/kWh}$} \\ \hline
    \multirow{1}{*}{Lines}  & -             & 47 nodes DN & 23.30/km & 20 \\ \hline \hline
  \end{tabular}
\end{table}

The conditions for different cases are summarized in Table~\ref{tab:cases}. Cases A through D involve joint planning of PV-EV stations and the DN, with case A\cite{lu2021two} and case B\cite{gao2025distributionally} addressing the RO problem using the C\&CG and iC\&CG algorithms, respectively. Case C \cite{xiang2023distributionally} uses the iC\&CG algorithm to solve the DRO problem, while case D solves the SRO problem using the A-iC\&CG algorithm. Additionally, compared to case D, cases E and F explore different types of RCS. Cases A through C consider traditional DIU, where EVs are assumed to charge at their designated areas, creating uncertainty in EV charging load across different regions, without considering DDU.
\vspace{-0.5em}
\begin{table}[htbp]
\centering
\vspace{-0.5em}
\caption{Conditions for Different Case Studies}
\vspace{-0.5em}
\label{tab:cases}
\setlength{\tabcolsep}{5pt}
\renewcommand{\arraystretch}{1.2}
\begin{tabular*}{\linewidth}{@{\extracolsep{\fill}}ccccccc@{}}
\hline\hline
\textbf{Case No.} & \textbf{Case A} & \textbf{Case B} & \textbf{Case C} & \textbf{Case D} & \textbf{Case E} & \textbf{Case F} \\
\hline
EV     & $\times$ & $\times$ & $\times$ & $\times$ & $\surd$ & $\times$ \\
PV-EV   & $\surd$ & $\surd$ & $\surd$ & $\surd$ & $\times$ & $\times$ \\
PV-ESS-EV  & $\times$ & $\times$ & $\times$ & $\times$ & $\times$ & $\surd$ \\
RO        & $\surd$ & $\surd$ & $\times$ & $\times$ & $\times$ & $\times$ \\
DRO       & $\times$ & $\times$ & $\surd$ & $\times$ & $\times$ & $\times$ \\
SRO       & $\times$ & $\times$ & $\times$ & $\surd$ & $\surd$ & $\surd$ \\
C\&CG     & $\surd$ & $\times$ & $\times$ & $\times$ & $\times$ & $\times$ \\
i-C\&CG   & $\times$ & $\surd$ & $\surd$ & $\times$ & $\times$ & $\times$ \\
Ai-C\&CG  & $\times$ & $\times$ & $\times$ & $\surd$ & $\surd$ & $\surd$ \\
\hline\hline
\end{tabular*}
\end{table}
\vspace{-1.0em}
\subsection{Results and Convergence Analysis}
This section presents the planning results and their associated costs, along with an analysis of the impact of different types of RCS on voltage quality.
\vspace{-0.5em}
\begin{figure}[htbp]
    \centering
    \includegraphics[width=0.45\textwidth]{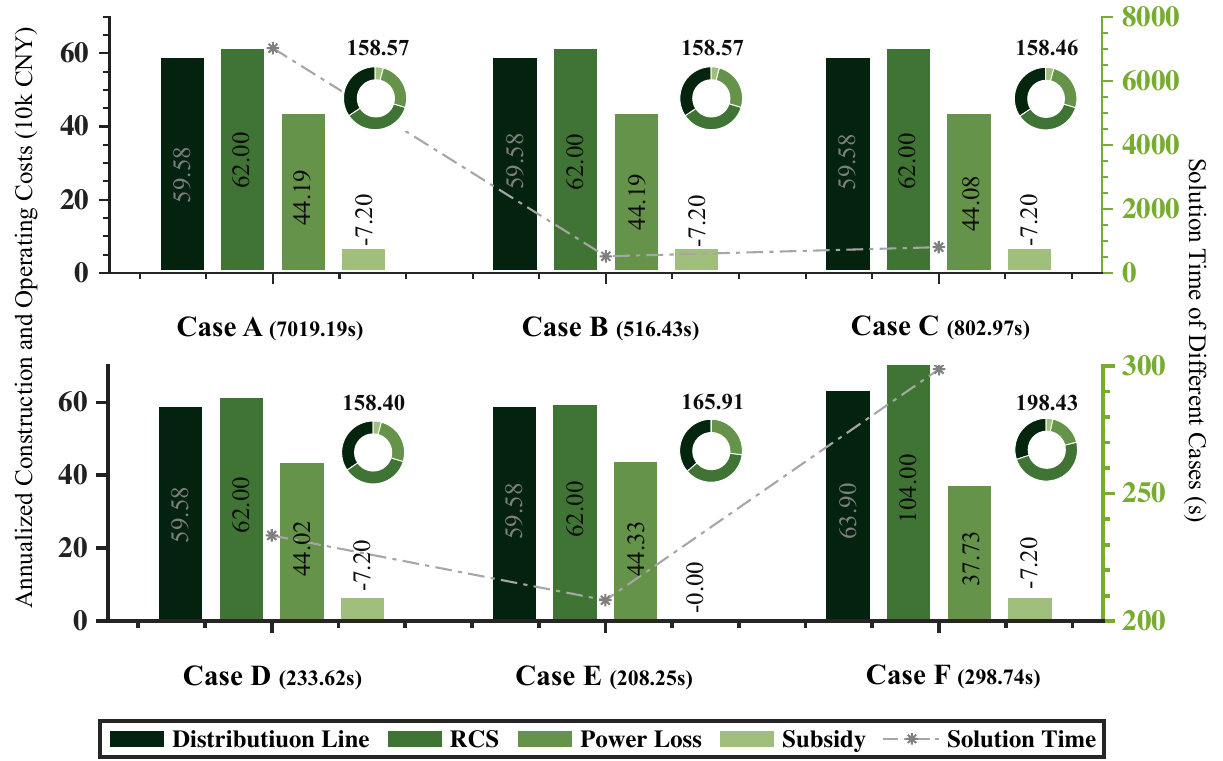}
    \vspace{-0.5em}
    \caption{Annualized costs ($10^{4}$ CNY \textyen) and solution time.}
    \label{fig6}
\end{figure}
\vspace{-0.5em}

Fig.~\ref{fig6} shows the annualized construction and operational costs, as well as the solution times. Convergence is considered achieved when the gap between the upper and lower bounds is less than or equal to 0.01\% (for C\&CG, \( UB^i \) and \( LB^i \); for iC\&CG and A-iC\&CG, \( \overline{UB} \) and \( LB^k \)). Compared to the C\&CG algorithm for RO (case A), the iC\&CG algorithm for RO (case B) significantly reduces the solution time from 7019.19s to 516.43s. Additionally, the iC\&CG algorithm for DRO (case C) obtains the optimal solution in 802.97s, while the proposed A-iC\&CG algorithm for SRO (case D) further reduces the solution time to 233.63s.
\vspace{-1.0em}
\begin{figure}[htbp]
    \centering
    \includegraphics[width=0.46\textwidth]{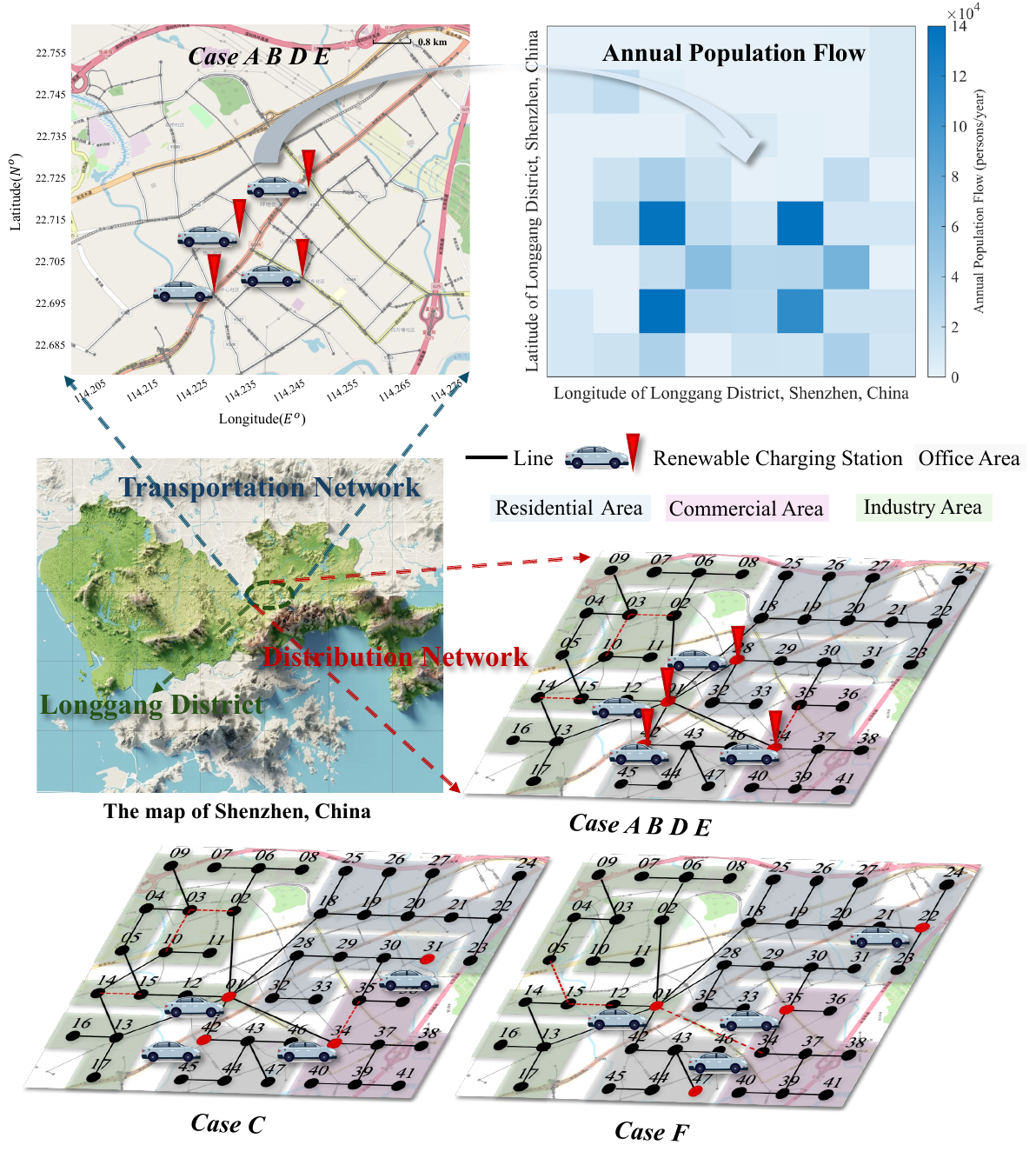}
    \vspace{-1.2em}
    \caption{Planning solutions and population flow in Longgang, Shenzhen, China.}
    \label{fig7}
\end{figure}
\vspace{-1.0em}

As shown in Fig.~\ref{fig7}, cases A, B, D, and E yield the same planning results for the DN and RCS siting. Since cases A and B solve the same RO problem, their costs are identical. Moreover, cases C and D do not consider the scenario where the number of EVs reaches its maximum, resulting in lower EV charging loads and consequently lower power losses. However, the solution time for case C is 3.44 times that of case D, highlighting the efficiency of the A-iC\&CG algorithm compared to both C\&CG and iC\&CG.
When comparing case D (PV-EV) with case E (EV-only) and case F (PV-EV-ESS), the costs are higher for both case E and case F. In case E, which lacks PV, there is a 0.71\% increase in power loss costs and a 4.74\% increase in total costs. Conversely, while the installation of ESS in case F reduces power loss by 14.29\%, it increases total costs by 25.27\% due to the high construction cost of ESS. These results suggest that the optimal RCS configuration in the coupled system is the PV-EV station.
In cases A, B, D, and E, the RCS are sited at nodes 1, 28, 34, and 42 in various locations, all of which are near substations to minimize power loss in the DN. Furthermore, the RCS sites at nodes 28, 34, and 42 in the transportation network are closely aligned with the annual population flow distribution, which corresponds to areas with higher population density. In contrast, the site at node 1 is less aligned with the population distribution but is strategically located near the substation. This indicates that siting decisions take into account not only the DN but also the transportation network and population distribution. 
However, in cases C and F, the RCS sites at node 31 (case C) and node 22 (case F) are located near the end of the DN. This suggests that the installation of PV and ESS in the RCS more effectively reduces power loss compared to EV charging, which tends to increase power loss. As a result, RCS are planned for locations at the end of the DN line to optimize power loss reduction.
\vspace{-1.0em}
\begin{figure}[htbp]
    \centering
    \includegraphics[width=0.45\textwidth]{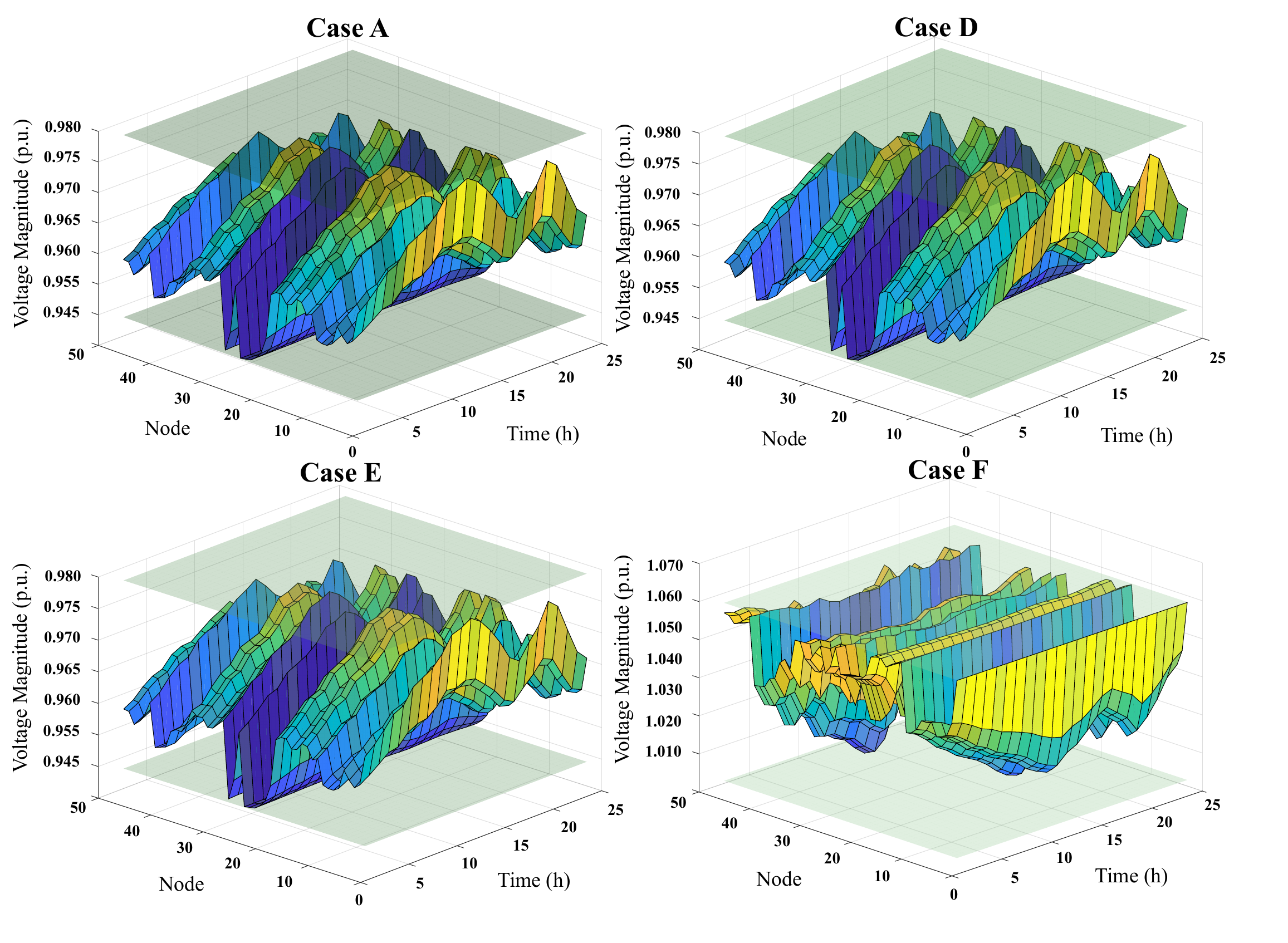}
    \vspace{-1.0em}
    \caption{Voltage distribution in different cases.}
    \label{fig8}
\end{figure}
\vspace{-1.0em}

Fig.~\ref{fig8} illustrates the voltage distribution for a typical day. Since cases A, D, and E share the same DN topology, their voltage profiles are similar, with an extreme difference of 0.057 p.u. and standard deviations of 0.01343 p.u., 0.01343 p.u., and 0.01340 p.u., respectively. The comparison between cases A and D shows that variations in the number of EVs within a reasonable range have minimal impact on the DN voltage. Comparing cases D and E reveals that the installation of PV reduces the standard deviation by 0.22\%. 
In contrast, the topology change in case F leads to a higher overall voltage. However, the extreme voltage difference (0.096 p.u.) and standard deviation (0.02313 p.u.) are significantly higher than those in the other cases, indicating that the change in topology results in poorer voltage quality.
\vspace{-1.0em}
\subsection{Sensitivity analysis of DDU and DIU}
This section analyzes how the EV charging load varies with the number of iterations and how the costs change with EV leet sizes and load fluctuation.
\vspace{-1.0em}
\begin{figure}[htbp]
    \centering
    \includegraphics[width=0.45\textwidth]{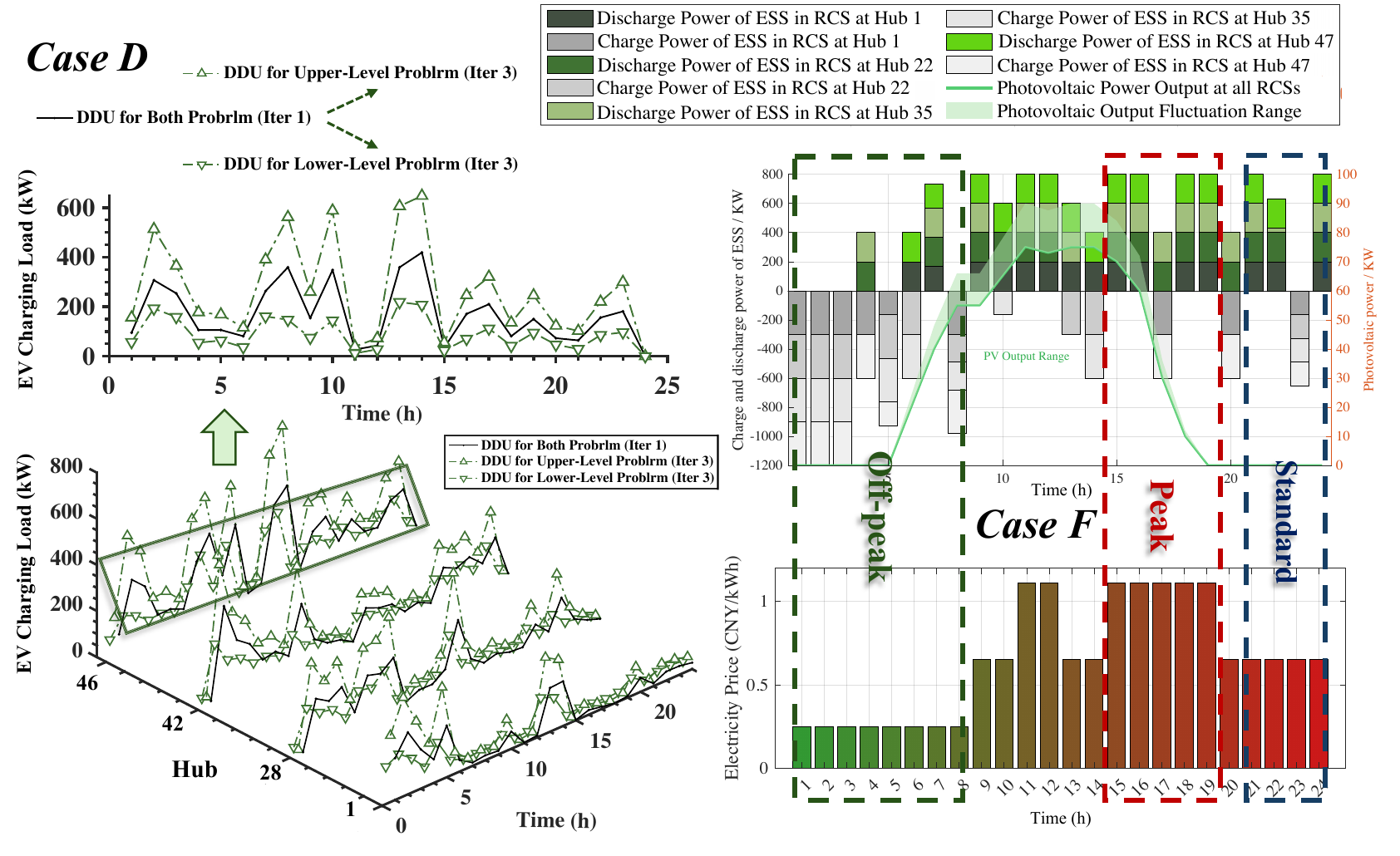}
    \vspace{-0.5em}
    \caption{DDU update with iterative process and operation of RCS.}
    \label{fig9}
\end{figure}
\vspace{-1.0em}

Fig.~\ref{fig9} shows the update of the EV charging load at hub 46 in case D on a typical day. As the number of iterations increases, the EV charging load in the upper-level problem increases, causing both \( UB^i \) and \( LB^i \) of the two-stage SRO problem to rise. Conversely, the EV charging load in the lower-level problem decreases, which reduces \( \overline{UB} \) in the two-stage SRO problem. This behavior results from the \emph{Adaptive Oracle Tracking} in the A-iC\&CG algorithm, ensuring the rapid convergence of the two-stage SRO problem.
The operation of the RCS in case F on a typical day is also shown in Fig.~\ref{fig9}. The TOU pricing is related to the load demand in the power system: during high load periods, the TOU price is high, and during low load periods, the TOU price is low. During off-peak hours (1-8 AM), most ESS units in the RCS charge from the grid, while during peak hours (15-19 PM), most ESS units discharge power back to the grid, smoothing the bimodal load curve. Additionally, the PV output remains at the minimum value for PV output fluctuation across all RCS, which increases the objective function. This illustrates how the algorithm captures the worst-case scenarios of DIU to achieve the optimal solution under the worst-case conditions.
\vspace{-1.0em}
\begin{figure}[htbp]
    \centering
    \includegraphics[width=0.45\textwidth]{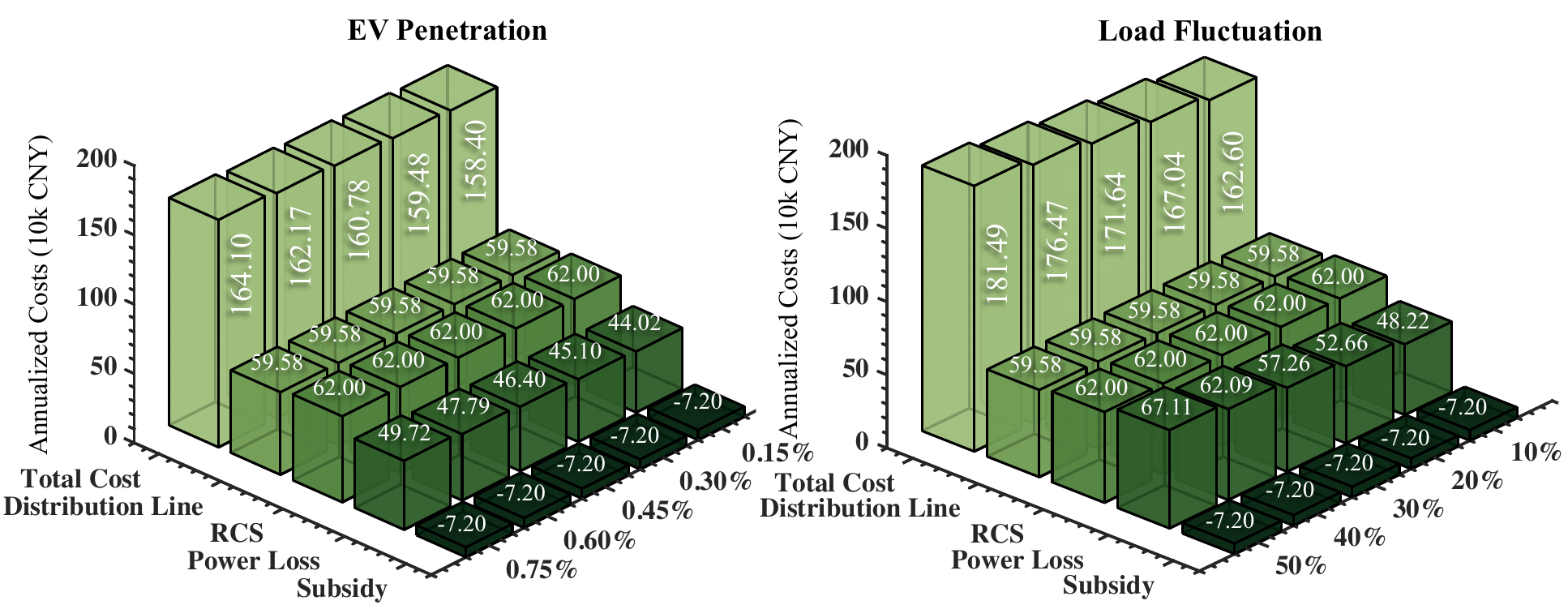}
    \vspace{-0.5em}
    \caption{Sensitivity analysis of EV penetration and load fluctuation.}
    \label{fig10}
\end{figure}
\vspace{-1.0em}

Fig.~\ref{fig10} shows how different planning costs change with EV penetration (as defined in equation (59) of reference \cite{wang2023joint}) and load fluctuation in case D. As EV penetration increases, both power loss and total costs increase. However, the planning results for the DN remain unchanged, indicating that the network topology is robust enough to handle the increase in EV penetration. This result further demonstrates the rationality of the SRO model, as small changes in EV penetration have minimal impact on DN planning. Since the number of EVs is random in the transportation system, it is reasonable to model this as DDU in the middle-level problem.
Similarly, as load fluctuation increases, power loss and total costs also increase. However, the planning results for the DN remain unchanged, suggesting that the network topology is sufficiently strong to accommodate load fluctuations.
\vspace{-1.0em}
\section{Conclusion}
\vspace{-0.3em}
This paper presented a tri-level, two-stage stochastic--robust co-planning framework that jointly determines distribution-network (DN) expansion and renewable charging station (RCS) siting while explicitly coupling transportation flows with power system operations. The model separates uncertainty into decision-dependent uncertainty (DDU) which captures EV-fleet size, routing, and charging behavior induced by siting decisions, and decision-independent uncertainty (DIU) in loads and PV generation at the operation stage. The lower-level operational problem is reformulated via KKT conditions, and storage complementarity is handled through an exact relaxation.
To solve the proposed large scale tri-level SRO efficiently, we developed an Adaptive inexact Column-and-Constraint Generation (A-iC\&CG) algorithm with an oracle-tracking mechanism for DDU. We proved finite iteration convergence and showed that the adaptive tightening of stochastic sets accelerates bound contraction across iterations.

Case studies on a 47-node DN coupled with 68-hub transportation network demonstrate three main insights. First, A-iC\&CG markedly reduces solution time compared with C\&CG and iC\&CG, while preserving optimality guarantees. Second, siting decisions concentrate near substations and high flow hubs, reflecting a strong co-location signal from both grid loss minimization and travel cost reduction. Third, among RCS configurations, the PV-EV option is cost optimal under the studied conditions. Voltage quality analyses further indicate that the proposed plan maintains acceptable profiles, and sensitivity studies show the network topology is robust to moderate EV penetration and load fluctuation changes.
\vspace{-1.0em}
\section{Appendix}
Table~\ref{RO} presents the iC\&CG algorithm for solving the two-stage RO problem. The parameter names are kept consistent with those in the A-iC\&CG algorithm.
\vspace{-0.5em}
\begin{table}[htbp]
\vspace{-0.5em}
\caption{I-C\&CG Algorithm for Solving Two-Stage RO}
\vspace{-0.5em}
\label{RO}
\scriptsize
\begin{tabular}{ll}
\hline \hline
\textbf{Step 1} 
& \textbf{Initialization:} Set $\overline{LB}=0$, $\overline{UB}=+\infty$, $i=1$, $k=1$, $\mathbb{U}=\emptyset$. 
\\ &Choose parameters $\mathbf{v}_l=\overline{v}$, $\varepsilon \in [0,1]$,$\tilde{\varepsilon} \in (0,\varepsilon/(\varepsilon+1))$, $\varepsilon^{i}_{up}$, \\ &$\alpha \in (0,1)$.\\
\textbf{Step 2} 
& \textbf{Upper-Level Problem:} \\
& Within an optimality gap $\varepsilon_{up}^{i}$, solve:
\\
& $\begin{array}{l}
\qquad \qquad \qquad \qquad \min_{\mathbf{y} \in \mathbb{Y}} \quad \mathbf{c^{\mathsf T}y} + \eta,\\
  \text{s.t. } 
  \begin{cases}
  \eta \ge \max_{\mathbf{u}_s, \mathbf{x}_{s,l}, \boldsymbol{\pi}_{s,l}, \boldsymbol{\lambda}_{s,l}, \mathbf{o}_{s,l}, \mathbf{w}_{s,l}}
  \tfrac{1}{2}\mathbf{x}_{s,l}^{\mathsf T}\mathbf{Q}_{s,l}\mathbf{x}_{s,l},\\
  \mathbf{c}^{\mathsf T}\mathbf{y} + \eta \ge \overline{LB}, \\
  \mathbf{A}\mathbf{y} \le \mathbf{b}, \\
  \text{Constraints (34)--(42).}
\end{cases}
\end{array} $    \\  
& Record the best feasible solution $(\mathbf{y}^i, \eta^i)$.
Record $LB^i \ge \overline{LB}$ and \\
& set $UB^i = \mathbf{c}^{\mathsf T}\mathbf{y}^i + \eta^i$.
  If $LB^i \ge \overline{LB}$, set $k=i$. \\
& Update $\overline{LB} = UB^i$.
\\
\textbf{Step 3} 
& \textbf{Lower-Level Problem:} \\
& For fixed $\mathbf{y}^i$, solve:
\\
& $\begin{array}{l}
\quad \max_{\mathbf{u}_s,\ \mathbf{x}_{s,l},\ \boldsymbol{\pi}_{s,l}, \boldsymbol{\lambda}_{s,l},\ \mathbf{o}_{s,l},\ \mathbf{w}_{s,l}}\frac{1}{2}\,\mathbf{x}_{s,l}^{\mathsf T}\mathbf{Q}_{s,l}\,\mathbf{x}_{s,l},\\
\qquad \qquad \qquad \text { s.t.} 
\quad \textnormal{Constraints (34)--(42)}. \\
\end{array} $    \\  
& Obtain the optimal $(\mathbf{x}_{s,l}^i,\mathbf{u}_s^i)$ with objective $D^i$.\\
& Update $\overline{UB} = \min\{\overline{UB}, \mathbf{c}^{\mathsf T}\mathbf{y}^i + D_i\}$.\\
\textbf{Step 4} 
& \textbf{Optimality Test and Backtracking:}  \\
& If $(\overline{UB}-LB^k)/\overline{UB}<\varepsilon$, then terminate and return $\mathbf{y}^i$; \\
& Otherwise, proceed as follows:\\
& \noindent $\bullet$ \textbf{Exploitation:} if $(\overline{UB} - UB^i)/\overline{UB} < \tilde{\varepsilon}$, set $i = k$, $\overline{LB} = LB^k$, \\
& and update $\varepsilon_{up}^{i} = \alpha \,\varepsilon_{up}^{i}$ for all $i \ge k$. Then go back to Step 2.\\
& \noindent $\bullet$ \textbf{Exploration:}  if $(\overline{UB}-UB^i)/\overline{UB}
\geq \tilde{\varepsilon}$, then go to Step 5.\\
\textbf{Step 5} 
& \textbf{Robust Scenario Set Enlargement:} \\
& Update $\mathbb{U} = \mathbb{U} \cup \{\,\mathbf{u}_{s}^i\}$.\\
& Set $i = i + 1$ and return to Step 2.\\
\hline \hline 
\end{tabular}
\end{table}
\vspace{-1.2em}
\bibliographystyle{IEEEtran}
\vspace{-0.2em}
\small\bibliography{reference}
\vspace{-1.0em}
\end{document}